\input harvmac
\noblackbox

\input epsf

\def\hat{\widehat}
\newcount\figno
\figno=0
\def\fig#1#2#3{
\par\begingroup\parindent=0pt\leftskip=1cm\rightskip=1cm\parindent=0pt
\baselineskip=11pt
\global\advance\figno by 1
\midinsert
\epsfxsize=#3
\centerline{\epsfbox{#2}}
\vskip 12pt
{\bf Fig.\ \the\figno: } #1\par
\endinsert\endgroup\par
}
\def\figlabel#1{\xdef#1{\the\figno}}
\def\encadremath#1{\vbox{\hrule\hbox{\vrule\kern8pt\vbox{\kern8pt
\hbox{$\displaystyle #1$}\kern8pt}
\kern8pt\vrule}\hrule}}

\def\np#1#2#3{Nucl. Phys. {\bf B#1} (#2) #3}

\def\half{{1\over 2}}

\def\b{{\beta}}

\def\a{{\alpha}}
\def\g{{\gamma}}
\def\D{{\Delta}}
\def\m{{\mu}}
\def\n{{\nu}}
\def\ep{{\epsilon}}
\def\d{{\delta}}
\def\o{{\omega}}
\def\G{{\Gamma}}
\def\ph{{\phi}}
\def\t{{\theta}}

\def\ch{{\chi}}

\def\r{{\rightarrow}}

\def\frac#1#2{{#1\over #2}}

\def\CN{{\cal N}}

\def\CL{{\cal L}}

\def\CR{{\cal R}}
\def\L{{\Lambda}}
\def\T{{\Theta}}

\def\p{\partial}
\def\cl{\centerline}

\lref\rsw{N. Seiberg and E.Witten, ``String Theory and Noncommutative
Geometry,'' hep-th/9908142, JHEP { \bf 09} (1999) 032.}
\lref\rcds{A. Connes, M. Douglas and A. Schwarz, ``Noncommutative 
Geometry and Matrix Theory: Compactification on Tori,'' hep-th/9711162, 
JHEP { \bf 02} (1998) 003.}

\Title
{\vbox{\baselineskip12pt
\hbox{PUPT-1905 IASSNS-HEP-99-112 hep-th/9912072}}}
{\vbox{\centerline{Noncommutative Perturbative Dynamics}}}

\centerline{Shiraz Minwalla, Mark Van
Raamsdonk\footnote{$^1$}{minwalla, mav@princeton.edu},}
\smallskip
\centerline{\sl Department of Physics, Princeton University}
\centerline{\sl Princeton, NJ 08544, USA}
\medskip
\centerline{and}
\medskip
\centerline{Nathan Seiberg\footnote{$^2$}{seiberg@sns.ias.edu}}
\smallskip
\centerline{\sl School of Natural Sciences, Institute for Advanced
Study}
\centerline{\sl Olden Lane, Princeton, NJ 08540, USA}

\vskip 0.8cm

\centerline{\bf Abstract}
\medskip
\noindent
We study the perturbative dynamics of noncommutative field theories on
$\CR^d$, and find an intriguing mixing of the UV and the IR.  High
energies of virtual particles in loops produce non-analyticity at low
momentum.  Consequently, the low energy effective action is singular
at zero momentum even when the original noncommutative field theory is
massive.  Some of the nonplanar diagrams of these theories are
divergent, but we interpret these divergences as IR divergences and
deal with them accordingly.  We explain how this UV/IR mixing arises
{}from the underlying noncommutativity.  This phenomenon is
reminiscent of the channel duality of the double twist diagram in open
string theory.

\vskip 0.5cm
\Date{Dec. 1999}

\newsec{Introduction}

\nref\rfilk{T.~Filk, ``Divergences in a Field Theory on Quantum
Space,'' Phys. Lett. {\bf B376} (1996) 53.}%
\nref\onevgb{J.C.~Varilly and J.M.~Gracia-Bondia, ``On the ultraviolet
behavior of quantum fields over noncommutative manifolds,''
Int.\ J.\ Mod.\ Phys.\ {\bf A14} (1999) 1305, hep-th/9804001.}
\nref\two{M.~Chaichian, A.~Demichev and P.~Presnajder, ``Quantum Field
Theory on Noncommutative Space-times and the Persistence of
Ultraviolet Divergences,'' hep-th/9812180;  ``Quantum Field Theory on
the Noncommutative Plane with E(q)(2) Symmetry,'' hep-th/9904132.}%
\nref\rjabbari{ M. Sheikh-Jabbari, ``One Loop Renormalizability
of Supersymmetric Yang-Mills Theories on Noncommutative Torus,''
hep-th/9903107,JHEP {\bf 06} (1999) 015.}%
\nref\rruiz{C.P.Martin, D. Sanchez-Ruiz,
``The One-loop UV Divergent Structure of U(1) Yang-Mills
 Theory on Noncommutative $R^4$,''hep-th/9903077,
Phys.Rev.Lett. {\bf 83} (1999) 476-479.}%
\nref\rwulkenhaar{T. Krajewski, R. Wulkenhaar, ``Perturbative quantum
gauge fields on the noncommutative torus,'' hep-th/9903187.}%
\nref\twopfo{S.~Cho, R.~Hinterding, J.~Madore and H.~Steinacker,
``Finite Field Theory on Noncommutative Geometries,''
hep-th/9903239.}%
\nref\three{E.~Hawkins, ``Noncommutative Regularization for the
Practical Man,'' hep-th/9908052.}%
\nref\rsusskind{D. Bigatti and L. Susskind, 
``Magnetic fields, branes and noncommutative geometry,'' 
hep-th/9908056.}%
\nref\rishibashi{N. Ishibashi, S. Iso, H. Kawai and Y. Kitazawa,
``Wilson Loops in Noncommutative Yang-Mills,'' hep-th/9910004.}%
\nref\riouri{ I. Chepelev and  R. Roiban, 
``Renormalization of Quantum Field Theories on Noncommutative $R^d$, I. 
Scalars, ''hep-th/9911098.}%
\nref\rbenaoum{ H.Benaoum, ``Perturbative BF-Yang-Mills theory 
on noncommutative $R^4$, ''hep-th/9912036.}%

In this note we follow \refs{\rfilk-\rbenaoum} and analyze quantum
field theory on a noncommutative space.  For simplicity we focus on
the case of the noncommutative $\CR^d$ and in most of the note we
discuss only scalar field theories (we will briefly mention gauge
theories in the last section).  Although noncommutative gauge theories
appear in string theory \rcds\ (see \rsw\ and references therein for
recent developments), through most of the paper our discussion will be
field theoretic.

The underlying $\CR^d$ is labeled by $d$ noncommuting coordinates
satisfying
\eqn\comm{ [x^{\m}, x^{\n}]= i \T^{\m\n}.}
Here $\T^{\m\n}$ is real and antisymmetric.  
By a choice of coordinates $\T$ can be brought to the form
\eqn\defchots{
\pmatrix{ 0 & \t_1 & \cr
       -\t_1 & 0 & \cr
    & & \ddots  }.}
Thus a $\T$ matrix of rank $r$ describes a spacetime with ${r\over 2}$
pairs of noncommuting coordinates and $d-r$ coordinates that commute
with all others.

The algebra of functions on noncommutative $\CR^d$ can be viewed as an
algebra of ordinary functions on the usual $\CR^d$ with the product
deformed to the noncommutative, associative star product, defined by
\eqn\starp{
\big( \ph_1 \star \ph_2 \big) (x)=e^{{i\over 2} \T^{\mu \nu} \p^y_\mu
\p^z_\nu} \ph_1(y) \ph_2(z)|_{y=z=x} \; .}
Thus, we will study theories whose fields are functions on ordinary
$\CR^d$, with actions of the usual form $S=\int d^d x {\cal
L}[\phi]$, except that the fields in ${\cal L}$ are multiplied using
the star product.

In section 2 we take a first look at the perturbation theory of
noncommutative scalar field theories.  After deriving the
Feynman rules, we review the results of Filk \rfilk, who showed that
planar diagrams in noncommutative field theories 
are essentially the same as those in the corresponding
commutative theory. 

Section 3 is devoted to a detailed analysis of one loop diagrams and
especially to nonplanar one loop diagrams in $\phi^3$ in six
dimensions and $\phi^4 $ in four dimensions.  These diagrams are UV
finite but exhibit interesting IR singularities.  In particular, even
though the theory we start with is massive, its correlation functions
exhibit singularities at zero momenta!

This nontrivial mixture between UV and IR phenomena is perhaps the
most surprising result of this paper.  It is tempting to speculate
that such mixing of scales could be relevant to various hierarchy
problems like the problem of the cosmological constant.

In section 4 we point out that the IR singularities of section 3 are
quite similar to the appearance of closed string poles in the double
twist diagram in open string theory.  It is surprising that a field
theory exhibits such a stringy phenomenon.

In section 5 we explore higher order Feynman diagrams and some of
their properties.  We also consider the limit of maximal
noncommutativity, $\T\to\infty$. In this limit the theory is dominated
by planar graphs and appears to be stringy.

In section 6 we explain how the noncommutativity of spacetime leads to
the surprising IR phenomena discussed in earlier sections.  Section 7
is the beginning of an investigation into the properties of
noncommutative gauge theories.  In an appendix we present the
expression for the Feynman integral corresponding to an arbitrary
graph in a noncommutative theory, written in terms of Schwinger
parameters.

\newsec{First Look at Perturbation Theory}

\subsec{Feynman Rules}

For any noncommutative theory, the quadratic part of the action is the
same as in the commutative theory, since
\eqn\quadac{\eqalign{
&\int d^d x  \ph \star \ph =\int d^dx \ph \ph\cr
&\int d^d x \partial \ph \star \partial \ph 
=\int d^dx \partial \ph \partial \ph\cr}}
(we have dropped total derivatives assuming suitable boundary conditions 
on $\phi$).  Therefore propagators take their usual form, as in
commutative theories.

The interactions are modified.  We consider a polynomial interaction
(perhaps with derivatives)
\eqn\intphi{\sum_{n=3}^L a_n g^{n-2} \phi^n,}
where the powers of the coupling constant $g$ are introduced for
convenience, the coefficients $a_n$ are arbitrary numbers, and
the products of $\phi$ are star products.  In momentum space the
interaction vertex of $\phi^n$ has an additional phase factor relative
to the commutative theory
\eqn\verfac{V(k_1, k_2..., k_n)=e^{-{i\over 2} \sum_{i<j} k_i \times
k_j,}} 
where $k_i$ is the momentum flowing into the vertex through the $i$th
$\ph$ and  
\eqn\kitimkj{k_i \times k_j \equiv k_{i\m} \T^{\m\n}k_{j\n}.}
This is the only modification to the Feynman rules.

$V$ is not invariant under arbitrary permutations of $k_i$ and so one
must keep track of the order in which lines emanate from vertices in a
Feynman diagram.  (Using momentum conservation, it is easy to see that
$V(k_1, \cdots ,k_n)$ is invariant under cyclic permutations of
$k_i$.) One way to keep track of the order of the lines is to view
$\phi$ as an $N\times N$ matrix and to use double line notation in the
Feynman diagrams.  In the noncommutative theory this is useful even
when $N=1$.  Contractions which yield distinct diagrams in double line
notation are associated with different contributions from the
noncommutative phases.

One may associate a genus with every diagram.  Specifically, the genus
of a diagram is the minimum genus of a surface upon which the diagram
may be drawn without any intersections.  In the large $N$ expansion,
graphs are organized according to their genus.  We will find it
convenient to use the same organization even for $N=1$.  In the next
subsection we will study the genus zero, planar diagrams.

\subsec{Planar Graphs}

In a planar graph it is always possible to regard momentum as an
additional index flowing along the double lines. That is, given an $L$
loop planar graph, the momenta of all lines in the graph may be
written in terms of  `momenta' $l_1, \cdots ,l_{L+1}$ that flow
unchanged along an index line (see Fig.\ 1) of the graph.  Like
fundamental $U(N)$ matrix indices, the index line momenta along
adjacent edges of a given propagator flow in opposite directions.
Therefore, the momentum through any propagator (or external line) in
the graph is given by $l_i -l_j$, where $l_i$ and $l_j$ are the index
line momenta that flow along the adjacent edges of the propagator.
Since index line momenta are always conserved, this construction
automatically implements momentum conservation at vertices.  Note that
such a construction is not possible in nonplanar graphs.

For any vertex of the graph, let the momenta entering the vertex
through the $n$ propagators be $k_1, k_2, \cdots, k_n,$ in cyclic
order. Then $k_1=l_{i_1}-l_{i_2}$, $k_2=l_{i_2}-l_{i_3}, \cdots
,k_n=l_{i_n}-l_{i_1}$, in terms of which $(\sum_{i<j}k_i \times
k_j)=l_{i_1}\times l_{i_2} +l_{i_2} \times l_{i_3} + \cdots + l_{i_n}
\times l_{i_1}$. Thus the phase factor at any interaction point may be
expressed 
as the product of $n$ terms, one for each incoming propagator,
\eqn\incopro{V= \prod_{j=1}^{n}e^{-{i\over 2}(l_{i_j} \times
l_{i_{j+1}})}.}
\fig{A planar graph in double line notation. The phase associated
with, say, the middle propagator is $l_2 \times l_3$ from the
vertex $V_1$ and -$l_2\times l_3$ from the vertex $V_2$, adding to
zero.} {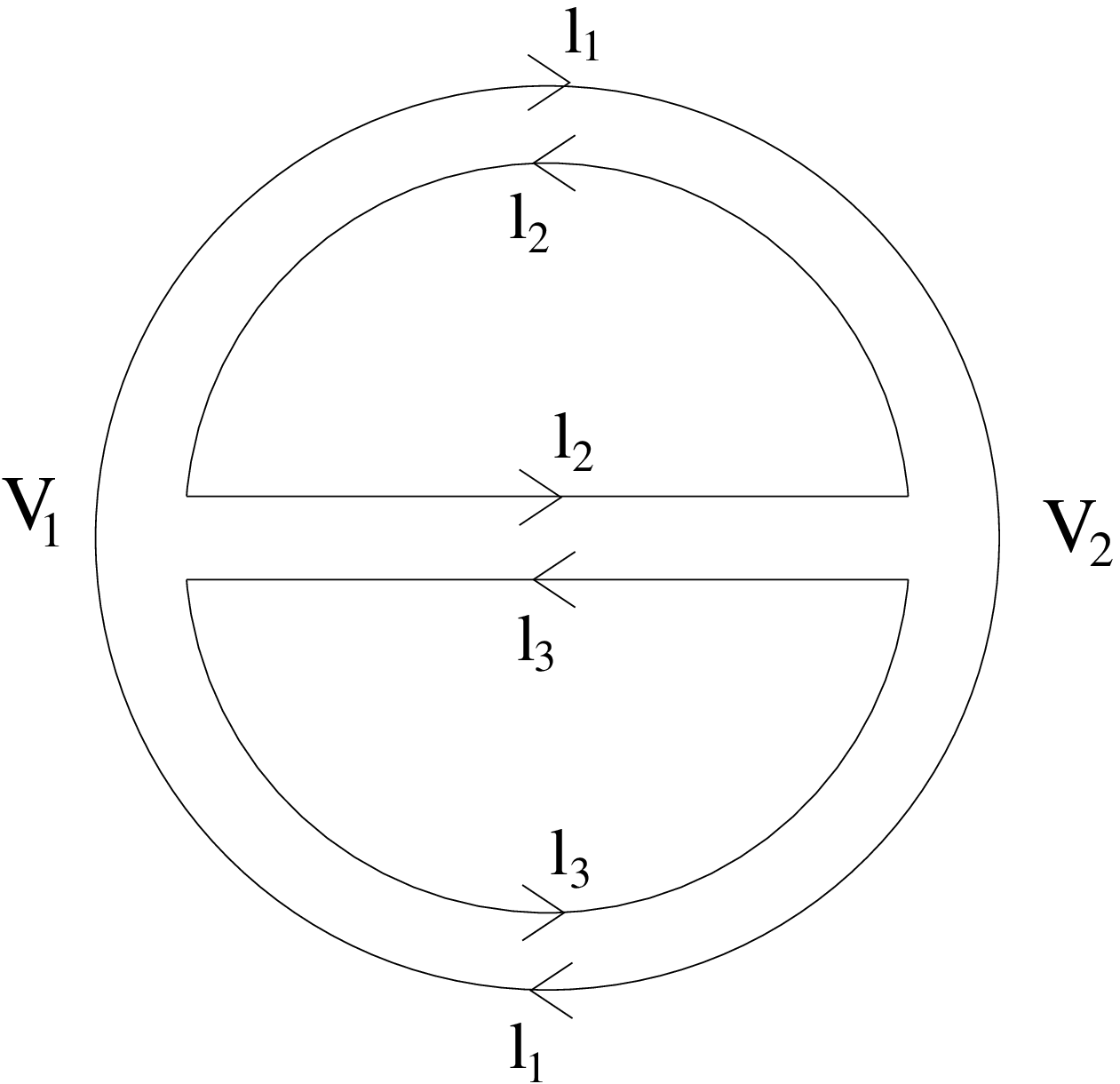}{2.0truein}
We see that the phase associated with any internal propagator is equal
and opposite at its two end vertices, and so cancels.  We conclude
that the phase factor associated with the planar diagram is
\eqn\phasefa{V(p_1, p_2..., p_n)=e^{-{i\over 2} \sum_{i<j} p_i \times
p_j},}
where the sum is taken over all external momenta (in the original
variables) in the correct cyclic order.  This result is originally due
to \rfilk\ and this derivation follows \rishibashi.  We note that this
phase factor is exactly the one found in \rsw\ by computing disk
amplitudes in string theory.

It is important that this phase factor is independent of the details
of the internal structure of the graph.  Thus, the contribution of a
planar graph to the noncommutative effective action is precisely the
contribution of the same graph in the $\T=0$ theory multiplied by
$V(p_1, p_2,..., p_n)$.  Such a $\T$ dependent phase factor is present
in all interaction terms in the bare Lagrangian and in all tree graphs
computed either with the bare Lagrangian or with the effective 
Lagrangian computed with planar graphs.

At $\T=0$, divergent terms in the effective action are products of
local fields.  $ V(p_1, p_2,...,p_n)$ modifies these to the star
product of local fields.  Divergences (due to planar graphs) in the
effective action at $\T \neq 0$ may therefore be absorbed into
redefinitions of bare parameters, if and only if the $\T=0$ theory is
renormalizable.

We stress that this renormalization procedure is not obtained by
adding local counterterms.  The added counterterms are of the same
form as the original terms in the Lagrangian but they are not local.

\subsec{Nonplanar Graphs}

Nonplanar diagrams have propagators that cross over each other, or
over external lines as in Fig 2.
\fig{Two lines crossing in a nonplanar graph.}{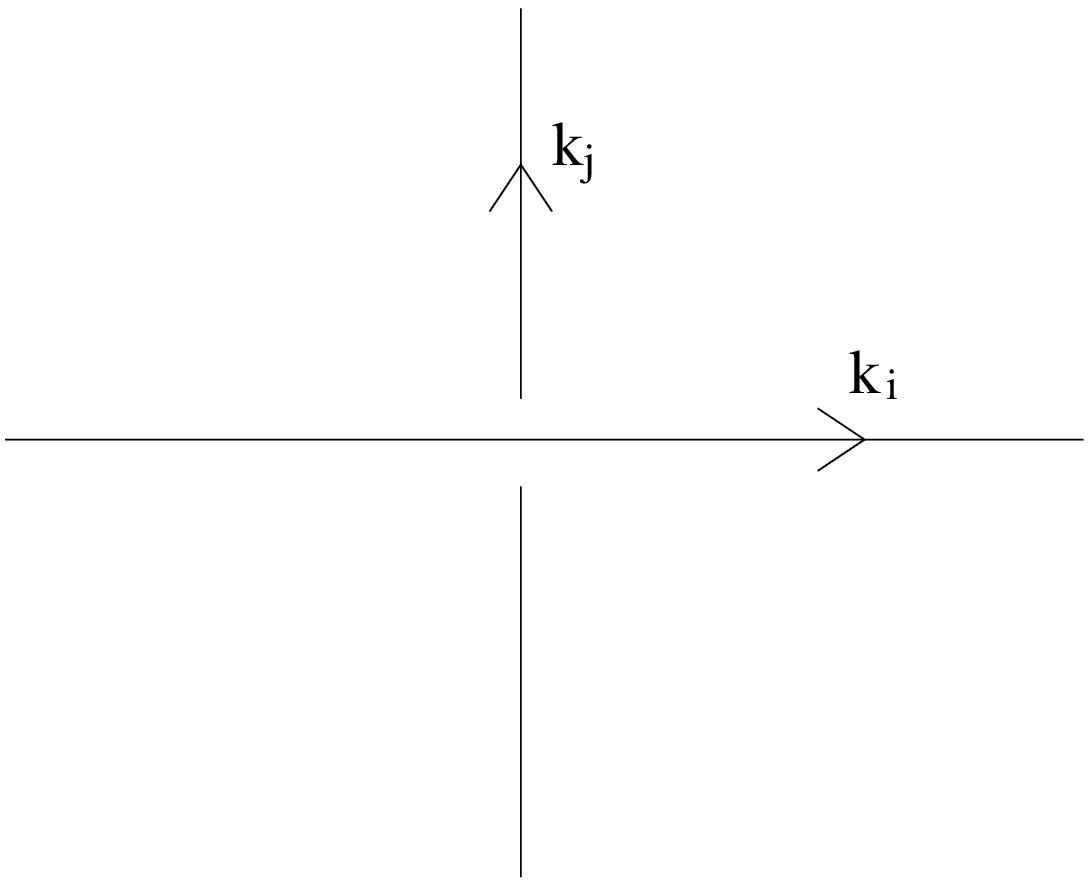}{2.0truein}
If, instead of crossing, the two lines in Fig.\ 2 had met at a (4 point)
vertex, the graph would have had an additional phase factor of
$e^{-{i\over 2}(k_j \times k_i -k_j\times k_i -k_i\times k_j + k_j
\times k_i)}=e^{-ik_j \times k_i}$ but would then have been planar (as
far as this crossing was concerned). Therefore, any nonplanar graph
will have an extra phase
\eqn\extraphase{e^{+i k_j \times k_i}}
for each crossing of momenta $k_i$ and $k_j$ in addition to the phase 
associated with the ordering of external momenta. The complete phase for 
a general graph may be written \rfilk\
\eqn\comppha{V(p_1, p_2,...,p_n) \; e^{-i\big( \half \sum_{i,j}C_{ij}
k_i\times k_j  \big) },}
where $V$ is as in \phasefa, and $C_{ij}$, the intersection matrix,
counts the number of times 
the $i^{th}$ (internal or external) line crosses over the $j^{th}$
line. Crossings are counted as positive if $k_i$ crosses $k_j$ with
$k_j$ moving to the left.

\fig{The overlaps in a nonplanar Feynman diagram can be chosen in
several different ways.}{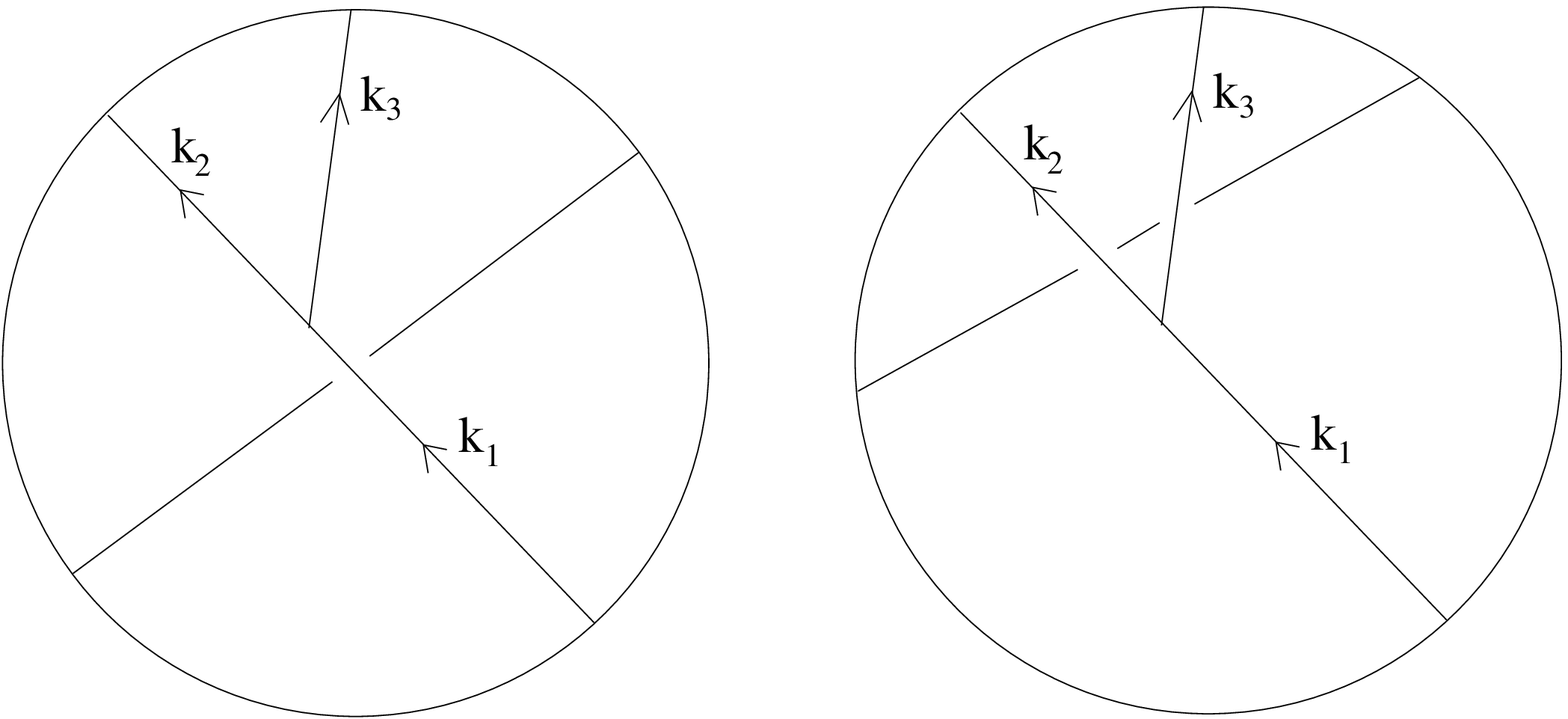}{3.0truein}
The matrix $C_{ij}$ corresponding to a given graph is not unique,
since different ways of drawing the graph will lead to different
intersections (see Fig.\ 3). However, all of these yield identical
Feynman integrands; the ambiguity corresponds to the fact that the
internal momenta $k_i$ of the graph are not all independent but are
constrained by momentum conservation at each vertex.

As the $\T$ dependence of nonplanar graphs does not factor out of the
integral, nonplanar graphs at $\T \neq 0$ behave very differently from
their $\T=0$ counterparts. For instance, as we will see in section 3,
all nonplanar one loop diagrams are finite.  The improved convergence
of nonplanar graphs is a result of the damping effects of rapid
oscillations of internal momentum dependent phase factors in the
integrand.  Since each nonplanarity in a Feynman diagram results in a
new internal momentum dependent phase factor in the Feynman integrand,
one is tempted to guess that every nonplanar graph is convergent.
More precisely, one might conjecture that, with the exception of
divergent planar subgraphs, there are no new divergences associated
with nonplanar graphs.  This would imply that, after the planar graphs
have been renormalized, no further renormalization is needed.

In fact, it turns out that nonplanar graphs (with no divergent planar
subgraphs) are not all finite in theories with quadratic or higher
divergences (including all scalar theories), as we will see in section
5. This would appear to threaten renormalizability of these theories,
since the nonplanar counterterms that cancelled these divergences
would be complicated functions of $\T$ and external momenta, and have
a very nonlocal and complicated form in position space. At higher
orders such counterterms would generate new divergences of
increasingly complicated form.  It seems unlikely that this process
would terminate with a finite number of terms written in terms of
star products.  However, we will interpret these divergences as IR
rather than UV divergences and will suggest a procedure to deal with
them without introducing counterterms.

\newsec{One loop in scalar field theory}

In this section we explicitly compute several one loop nonplanar graphs
in $\ph^4$ theory in four dimensions and $\ph^3$ theory in six
dimensions.  We find that one loop nonplanar graphs in noncommutative
field theories are convergent at generic values of external momenta.
This is a consequence of the rapid oscillations of the phase factor
$e^{ip\times k}$ where $p$ is an external momentum and $k$ is the loop
momentum.  As this phase factor is zero when $p_{\m}\T^{\m\n}$ vanishes
(i.e.\ when either $\T$ or $p_{nc}$, the projection of $p$ onto the
noncommutative subspace, vanishes), the nonplanar graph is singular at
small $|p_{\m}\T^{\m\n}|$.  Indeed, the effective cutoff for a one loop
graph in momentum space is ${1\over \sqrt{-p_{\m} \T^2_{\m\n} p_{\n}}}$
where $p$ is some combination of the external momenta in the process.
Therefore turning on $\T$ replaces the UV divergence with singular IR
behavior.  This effect has interesting dynamical consequences for
noncommutative field theories, some of which are explored below.

\subsec{Quadratic Effective action in $\phi^4$ theory in $d=4$}

We begin with $\phi^4$ theory in four dimensions with the Euclidean
action 
\eqn\eucaction{S = \int d^4 x \big( {1 \over 2} (\partial_\mu \phi)^2
+ \half  m^2 \phi^2 + {1 \over 4!} g^2 \phi \star \phi \star \phi
\star \phi \big) . }
Consider the $1PI$ two point function, which at lowest order is simply
the inverse propagator
\eqn\lowpro{\Gamma^{(2)}_0 = p^2 + m^2.}
In the noncommutative theory, this receives corrections at one loop
{}from the two diagrams of Fig.\ 4, one planar and the other
non-planar. 
\fig{Planar and nonplanar one loop corrections to $\Gamma^{(2)}$ in
$\phi^4$ theory.}{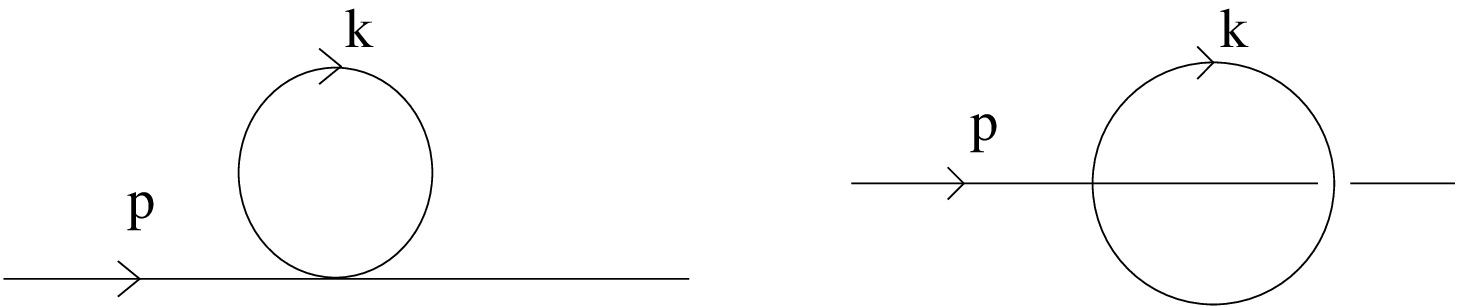}{3.0truein}
\noindent
The two diagrams (which are identical in the $\T=0$ theory up to a
symmetry factor) give
\eqn\twodia{\eqalign{
\Gamma^{(2)}_{1 \; planar} &={g^2 \over 3(2\pi)^4}
\int {d^4 k \over k^2 +m^2} \cr
\Gamma^{(2)}_{1 \; nonplanar} &={g^2 \over 6(2\pi)^4}
\int {d^4 k \over k^2 +m^2}
e^{ik \times p} \cr}}

The planar diagram is proportional to the one loop mass correction of
the commutative theory, and is quadratically divergent at high
energies.  In order to see the effect of the phase factor in the
second integral we rewrite the expressions for the two integrals in
terms of Schwinger parameters
\eqn\schwiner{{1 \over k^2+ m^2 } = \int_0^\infty d \alpha \;
e^{-\alpha (k^2 + m^2)} . }
The $k$ integrals are now Gaussian, and may be evaluated to yield
\eqn\kintegral{\eqalign{
\Gamma^{(2)}_{1 \; planar} &= {g^2 \over 48 \pi^2} \int {d\a
\over\a^2} e^{-\a m^2}\cr
\Gamma^{(2)}_{1 \; nonplanar} & ={g^2 \over 96 \pi^2}
\int {d\a \over\a^2} e^{-\a m^2 - {p \circ p \over \alpha} } ,\cr}}
where we have introduced new notation
\eqn\notat{p \circ q \equiv
-p_{\m}\T^2_{\m\n}q_{\n}=|p_{\m}\T^2_{\m\n}q_{\n}|}
(note that $p \circ p$ has dimension of length squared).  In order to
regulate the small $\a$ divergence in \kintegral\
we multiply the integrands in the
expressions above by $exp(-1/(\Lambda^2 \alpha))$ to get
\eqn\planint{\eqalign{ \Gamma^{(2)}_{1 \; planar} &= {g^2 \over 48 
\pi^2} \int {d\a \over\a^2} e^{-\a m^2 -{1\over \L^2 \a}}\cr
\Gamma^{(2)}_{1 \; nonplanar} & ={g^2 \over 96 \pi^2}
\int {d\a \over\a^2} e^{-\a m^2 - {p \circ p +{1\over \L^2} \over
\alpha} }. \cr}}

\noindent
Therefore,
\eqn\plancut{ \eqalign{
\Gamma^{(2)}_{1 \; planar} & = {g^2 \over 48 \pi^2}
\big(  \Lambda^2 -  m^2 {\rm ln} (
{\Lambda^2 \over m^2}) + O(1) \big) \cr
\Gamma^{(2)}_{1 \; nonplanar} & = {g^2 \over 96 \pi^2}
\big(  \Lambda_{eff}^2 -  m^2 {\rm ln} (
{\Lambda_{eff}^2 \over m^2}) + O(1) \big),  \cr }}
where 
\eqn\lambdaeff{\Lambda_{eff}^2 ={ 1 \over 1/\Lambda^2 + p \circ p }.}
 
In the limit $\L \r \infty$, the nonplanar one loop graph remains
finite, effectively regulated by the noncommutativity of
spacetime. In this limit  the effective cutoff $\L^2_{eff} ={1\over p
\circ p}$ goes to infinity when either $\T \r 0$ or $p_{nc} \r 0$.

The one loop 1PI quadratic effective action is
\eqn\effectiveaction{\eqalign{
S^{(2)}_{1PI}=\int d^4 p & \half \Big( p^2 + M^2 +
{ g^2  \over 96 \pi^2(p \circ p+{1\over \L^2})} - {g^2 M^2 \over 96  
\pi^2}\ln \left( {1\over M^2(p \circ p +{1\over \L^2}) } \right)
+ \cr 
& \cdots + \CO(g^4) \Big)  \ph(p) \ph(-p) \cr }}
where $M^2 = m^2 + {g^2\L^2 \over 48 \pi^2} - {g^2 m^2 \over 48 \pi^2}  
\ln \left( {\L^2 \over m^2} \right)...$ is the renormalized
mass. Consider the two cases 
\item{a)} $p \circ p \ll {1 \over \L^2}$ , and in particular the 
zero momentum limit. Here $\L_{eff} \approx \L$, and we recover
the effective action of the commutative theory,
\eqn\effaleffappl{S^{(2)}_{1PI}=
\int d^4 p  \half\left( p^2 + M'^2 \right) \ph(p) \ph(-p),}
where $M'^2=M^2+3{g^2 \L^2 \over 96 \pi^2} - {3g^2 m^2 \over 96 \pi^2} 
\ln \left( { \L^2 \over m^2} \right)...\ $.
If $M$ is fine tuned to be cutoff independent, then $M'$ and also
$S^{(2)}_{1PI}$ diverge as $\L \r \infty$.
\item{b)} $p \circ p  \gg {1 \over \L^2}$ and in particular the limit
$\L \r \infty$.  Here $\L^2_{eff} ={1\over p \circ p }$, and
\eqn\effectiveactionb{\eqalign{
S_{eff}=\int d^4 p  \half &\Big( p^2 + M^2 +
{ g^2 \over 96 \pi^2 p \circ p} \cr & - 
{g^2 M^2 \over 96 \pi^2} \ln\left( { 1\over m^2 p \circ p }\right)
+ ... +\ \CO(g^4) \Big) \ph(p) \ph(-p).\cr} }

The fact that the limit $\L \r \infty$ does not commute with the low
momentum limit $p_{nc} \r 0$ demonstrates the interesting mixing of
the UV $(\L \r 0)$ and IR ($p \r 0$) in this theory.  We will say more
about this below.

\subsec{Quadratic effective action in $\phi^3$ theory in $d=6$}

We repeat the computation performed above for $\phi^3$ theory in six
dimensions. Apart from a crucial sign, our results will mimic those of
the previous subsection.

Let
\eqn\phicuac{S = \int d^6 x \big( {1 \over 2} (\partial_\mu \phi)^2 +
\half m^2 \phi^2 + {g \over 3!} \phi \star \phi \star \phi \big) .}
As in $\phi^4$ theory, $\Gamma^{(2)}$ receives contributions both from a 
one loop planar
and a one loop nonplanar diagram (Fig.\ 5).
\fig{Planar and nonplanar one loop corrections to $\Gamma^{(2)}$ in
$\phi^3$ theory.}{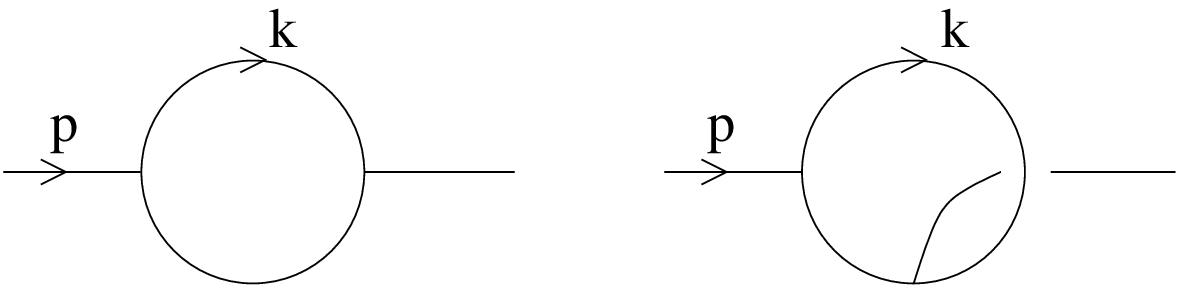}{3.0truein}
\noindent
The contribution of the nonplanar diagram may be written as
an integral over Schwinger parameters as
\eqn\gammatwoanon{\eqalign{
\Gamma^{(2)}_{1 \; nonplanar} &= - {g^2 \over 2^8 \pi^3}
\int { d\a_1 d\a_2 \over (\alpha_1 + \alpha_2)^3 }
e^{-m^2(\a_1+\a_2) - {p^2 \a_1\a_2
\over {\a_1 +\a_2}} - {p \circ p +{1\over \L^2}
\over \a_1 + \a_2}} \cr
\Gamma^{(2)}_{1 \; planar} &=\Gamma^{(2)}_{1 \; nonplanar}(\T=0) .\cr}}
The planar graph has a quadratic divergence from the region
where both $\alpha$'s are small. As above, this divergence is
effectively cutoff in the nonplanar graph, producing a ($\L \r
\infty$) quadratic effective action
\eqn\efff{S^{(2)}_{1PI}  = \int d^6 p {1 \over 2} \left( p^2+ M^2 -
{g^2 \over 2^8 \pi^2 p \circ p } + {g^2 \over 2^9 3 \pi^3}(p^2 +6M^2)
\ln({1 \over M^2  p \circ p })+... \right) \ph_{R}(p)
\ph_{R}(-p),} 
where $M$ is the planar renormalized mass, and $\ph_R$ the planar 
renormalized scalar field. 

\subsec{Validity of the 1-loop approximation}

Although nonplanar contributions to \effectiveactionb\ and \efff\ are
subleading in $g^2$, they are singular as $p_{nc} \r 0$, and so cannot
be ignored.  In particular, the nonplanar term significantly modifies
$S_{1PI}^{(2)}$ for $(p^2+M^2)p \circ p < \CO(g^2)$.

Since the singular dependence of $\Gamma^{(2)}_{ nonplanar}$ on $p
\circ p $ replaces the divergent dependence of $\G^{(2)}$ on $\L$ at
$\T=0$, the leading singularity in $\Gamma^{(2)}_{ nonplanar}$ at
$n^{th}$ order in perturbation theory is $g^{2n}{1 \over p \circ p }
\left( \ln(p \circ p M^2) \right)^{n-1}$ (we will study $n=2$ in section
5).  These higher order contributions are significant when compared to
the first order effect only for momenta such that $g^{2n}{1 \over
p \circ p } \left( \ln \left(p \circ p M^2 \right)\right)^{n-1}
\approx {g^2 \over p \circ p}$, i.e.\ for $ M^2 p \circ p \leq
\CO(e^{-{c\over g^2}})$ ($c$ is a positive constant).  Thus the 1-loop
approximation to $\Gamma^{(2)}_{nonplanar}$ is both important and
valid for
\eqn\impoanva{\CO(e^{-{c\over g^2}} )\leq M^2 (p \circ p) \leq
\CO(g^2).} 
The $p$ dependence can be trusted at low momentum except for
nonperturbatively small values of $p$.

\subsec{Stability of the Perturbative Vacuum: with and without 
Fermions}

Both \effectiveactionb\ and \efff\ take the form
\eqn\effect{S^{(2)}_{1PI} = \int d^6p {1 \over 2} \phi(p) \phi(-p)
(p^2 + M^2 +  h{g^2 \over p \circ p } + \{ {\rm subleading } \}).}
The constant $h$ is positive for the $\ph^4$ theory in $d=4$, but
negative for the 
$\ph^3$ theory in $d=6$, leading to a qualitative difference between
the dynamics of the two examples.

When $h$ is positive, the coefficient of $|\ph(p)|^2$ in \effect\ is
positive for all $p$.  The nonplanar one loop contribution to \effect\
modifies the $\ph$ propagator at small momenta, inducing long range
interactions (see the next subsection).

When $h$ is negative, the coefficient of $|\ph(p)|^2$ in \effect\ is
negative for momenta so small that $(p^2 + M^2){p \circ p} <\CO(g^2)$.
In order to minimize the effective action, $\ph(p)$ attains a vacuum
expectation value, at these low momenta.  In other words, the
perturbative vacuum is unstable.

Since the perturbative vacuum for $\ph^3$ theory is unstable anyway,
the observation of the previous paragraph may not seem too dramatic.
However, a similar effect occurs in a suitably modified $\ph^4$ theory
in four dimensions.  Recall that the term ${hg^2 \over p \circ p }$ in
\effect\ results from the quadratic divergence of the nonplanar
diagram of the commutative theory.  The sign of the infrared pole is
therefore correlated with the sign of the mass renormalization (due to
nonplanar graphs) of this theory.

Now adding sufficiently many (or sufficiently strongly) Yukawa coupled
fermions to the $\ph^4$ theory changes the sign of the mass
renormalization.  (Recall that a model with a single complex scalar
and a single chiral fermion with Yukawa coupling $g$ is
supersymmetric, has no quadratic divergences, and therefore $h=0$.)
Thus, in such a theory the sign of the $p\circ p$ leads to an
instability and the low energy theory has to be analyzed carefully.
We will return to this problem below.

\subsec{Poles in the Propagator}

The propagator derived from \effect\ has two poles.  The first is the
continuation to weak coupling of the zero coupling pole at
$p^2+m^2=0$, and occurs at 
\eqn\firstpole{p^2+m^2= \CO(g^2).}
It corresponds to the fundamental $\phi$ quanta.  The second is absent
at zero coupling and occurs at
\eqn\secondpole{p \circ p=-{g^2 h \over p_c^2 +m^2} + \CO(g^4),}
where $p_c$ is the restriction of $p$ to the commutative subspace 
($p_c=0$ if $\T$ is of maximal rank).

The second pole \secondpole\ formally signals the presence of a new
`particle' (we will qualify this below), whose mass goes to zero as
$g^2$ is taken to zero. This is true even though the theory is
massive, with mass $m$.  In fact, $m$ plays no significant role in the
new pole.

The appearance of the second pole in the $\ph$ propagator might
make one wonder if the nonlocality of the tree level noncommutative
action was a consequence of a massless particle having been integrated
out.  We do not believe this is the case.  Integrating out a massless
particle produces a nonlocal Lagrangian that is nonanalytic in momenta
around $p=0$, while the noncommutative action is completely analytic in
momenta at all $p$.  Indeed, the low momentum singularity in the
effective action results from integrating out very high (rather than
very low) momenta.  The light mode in the propagator is a consequence of
very high energy dynamics!

In Euclidean space, in the limit $g^2 \r 0$, both poles in the
propagator occur at imaginary values of momenta (in the case $h>0$).  In 
Lorentzian signature, we have to choose $\T^{0i}=0$.  Then $p^2+m^2=0$ 
can be satisfied with real momenta, but $p \circ p=0$ cannot (except for 
the trivial solution $p_{nc}=0$). Thus the
second pole corresponds to a Lagrange multiplier, rather than a
propagating particle.

At small finite values of $g^2$, $p \circ p=-{g^2 h \over p_c^2 +m^2}$
can be satisfied by real momenta in {\it Euclidean} space if and only
if $h$ is negative. A pole at real values of Euclidean momenta
corresponds to a tachyonic instability of the vacuum, in accord with
the discussion of the previous subsection. As discussed there, the
sign of $h$ depends on the details of the theory.

The pole in the propagator at small values of $p$ has dramatic
consequences.  In position space it leads to long range correlations.
Normally correlation functions decay exponentially, with the decay
constant given by the mass $m$.  In the noncommutative theory, with
$\T$ of maximum rank, they decay algebraically for small $g$, and the
small $g$ corrections lead to exponential decay, but with decay
constant of order $g$.  We should stress that all this is only for
positive $h$.  Otherwise, the theory is tachyonic and suffers from the
instability discussed above.

\subsec{Wilsonian Lagrangian}

Consider a Wilsonian action with a cutoff $\L$, of the form
\eqn\effact{S_{eff}( \L)
=\int d^4x {Z(\L) \over 2} \left( (\p \ph)^2 + m^2(\L) \ph^2 \right)
+ {g^2(\L)Z^2(\L) \over 4!} (\ph \star \ph\star \ph \star \ph).}
The statement that the noncommutative $\ph^4$ theory is renormalizable
would normally imply that it is possible to choose the functions
$Z(\L)$, $m(\L)$ and $g^2(\L)$ in such a way that
\item{a)} Correlation functions computed with this Lagrangian have a
limit as $\L \r \infty $.
\item{b)} Correlation functions computed at finite $\L$ differ from
their limiting values by terms of order ${1\over \L}$ for all values
of momenta.

Property b) is manifestly untrue of the noncommutative scalar theories
under consideration.  While it is presumably possible to choose
$Z(\L)$, $m(\L)$ and $g^2(\L)$ in such a way that the $\L \r \infty$
limit of all correlation functions exists, the various correlation
functions (at various values of momenta) do not converge uniformly to
their limiting values.  As we have seen above,  the two point function
computed using \effact\ at any finite value of $\L$ differs
significantly from its $\L \r \infty$ value for small enough
momentum $p_{nc}$ (for $p \circ p \L^2 \ll 1$).  In particular, as we
have noted previously, the limit $p_{nc} \r 0$ does not commute with
the limit $\L \r \infty$.

We attempt to find a substitute for \effact\ which correctly captures
the leading small momentum singularity in $\langle \ph \ph \rangle$ at
finite $\L$.  To this end we introduce a new degree of freedom into
the Wilsonian action, that reproduces the important low energy effects
arising from integrating out the modes between $\L$ and $\infty$.  The
new degree of freedom should give a contribution to the $\ph \ph$ two
point function that is small for $p \circ p \L^2 \gg 1$ and is
approximately ${1 \over p \circ p } $ for $p \circ p\L^2 \ll
1$. This is achieved by introducing a new field $\ch$ into the
modified Wilsonian action
\eqn\modact{S_{eff}'(\L)= S_{eff}(\L) + \int d^4 x \left(\half \p \ch
\circ \p \ch  +\half \L^2 \left( \p \circ \p \ch \right)^2 + i{1 \over
\sqrt{96  \pi^2}}g\ch \ph \right).} 
$\ch$ appears quadratically in $\modact$ and so may be integrated out, 
yielding
\eqn\modactb{S_{eff}'(\L)= S_{eff}(\L)+ \int {d^4 p \over 2}   \ph(p)
\ph(-p) {1 \over 96 \pi^2} \left({ g^2 \over p \circ p} -{g^2 \over p 
\circ p  +{1\over
\L^2}} \right).}
As we have seen above, $S_{eff}$ leads to the quadratic 1PI effective
action \effectiveaction\ at $\CO(g^2)$.  Therefore, the quadratic 1PI
effective action at $\CO(g^2)$ resulting from $S_{eff}'$ is
\eqn\effectiveactionc{S_{1PI}=\int {d^4 p \over 2}  \left( p^2 + M^2
+{ g^2 \over 96 \pi^2 p \circ p} + {g^2 M^2\over 96 \pi^2}\ln \left(
{1\over M^2(p \circ p+{1\over \L^2}) } \right) + \cdots \right) \ph(p)
\ph(-p)} 
correctly reproducing the leading low momentum singularity of the true
1PI effective action \effectiveactionb\ but incorrectly cutting off
the logarithmic singularity.

\subsec{Logarithms}

So far the effects we have been discussing arose from what started as
quadratic divergences in the commutative theory.  These turned into
poles, $1\over p\circ p$, in the noncommutative theory, which we
interpreted as the $\chi$ pole in the Wilsonian action.  Logarithmic
divergences are more common than quadratic divergences.  They occur
even when the quadratic divergences are absent as is the case in
supersymmetric theories or in gauge theories.  They also occur in the
scalar theories which we have been discussing.

The logarithmic UV divergences in the nonplanar graphs of the
commutative theory turn into terms of the form $\ln p\circ p$, which
are nonanalytic around $p=0$.  Rather than a pole, $1\over p\circ p$,
we now have a cut, $\ln p\circ p$, which is also an IR singularity.

Unfortunately, we do not fully understand the implications of the
logarithmic singularity in \effectiveactionb.  Consider the Wilsonian
effective action at small $\L$. The dynamics of $\ph$ freezes out at
these energies (as $\ph$ is massive), and so $\ph$ is effectively
classical. $\ch$ is a free light quantum field (its four derivative
kinetic term is negligible at low energies).  The low energy Wilsonian
effective action should correctly reproduce the low energy correlation
functions of $\ph$.  The effective action \modact\ (which is expected
to be correct for large $\L$) cannot accurately approximate the small
$\L$ effective action, because correlation functions computed using
\modact\ at small $\L$ correctly reproduce the pole but do not
reproduce the logarithm.  The correct small $\L$ effective action must
contain new dynamics, absent in \modact.  We see two possibilities for
the missing dynamics:
\item{a)} The classical field $\ph$ can couple to two massless
particles, resulting in a cut in its two point function.
\item{b)} The classical field $\ph$ can couple to a low energy conformal
field theory, with operators of dimensions of order $g^2$; the
logarithm represents the first term in the expansion of $\left( {p
\circ p } m^2 \right)^{cg^2+...}$.

\noindent
Clearly, these logarithms need better understanding.  We should point
out, however, that regardless of the details, just like the poles,
they demonstrate an IR singularity due to a UV phenomenon!

\subsec{Vertex corrections}

We end this section with a brief look at the contribution of one loop
nonplanar graphs to vertex corrections. The graphs we will compute are
logarithmically divergent in the commutative theory, and so will
depend logarithmically on $p^i\circ p^j $, where $p^i$ and $p^j$ are
external momenta.  Since we are primarily interested in low momentum
singularities, we will ignore all terms regular in $p^i$ at $p^i=0$. In 
particular, external momentum dependent phase factors that
factor out of the integral will be ignored.

\bigskip

\cl{\it Noncommutative $\ph^3$ in six dimensions}

Eight graphs contribute to the one loop vertex correction in $\ph^3$
theory in six dimensions. Two of these are planar, (e.g. the first
graph in Fig.\ 6), and differ in the cyclic ordering of external lines.
The remaining six graphs are nonplanar (e.g. the second graph in
Fig.\ 6), and differ in the choice of the external line that cuts an
internal line (3 choices), as well as the cyclic ordering of external
lines, given this choice (2 choices). Graphs that differ only in the
cyclic ordering of external lines differ only by an external momentum
dependent phase factor. For small external momenta, these phases are
unimportant, so  the one loop
vertex correction receives four distinct contributions, each with 
symmetry factor ${1\over 4}$.
\fig{Planar and nonplanar one loop corrections to $\Gamma^{(3)}$ in
$\phi^3$ theory.}{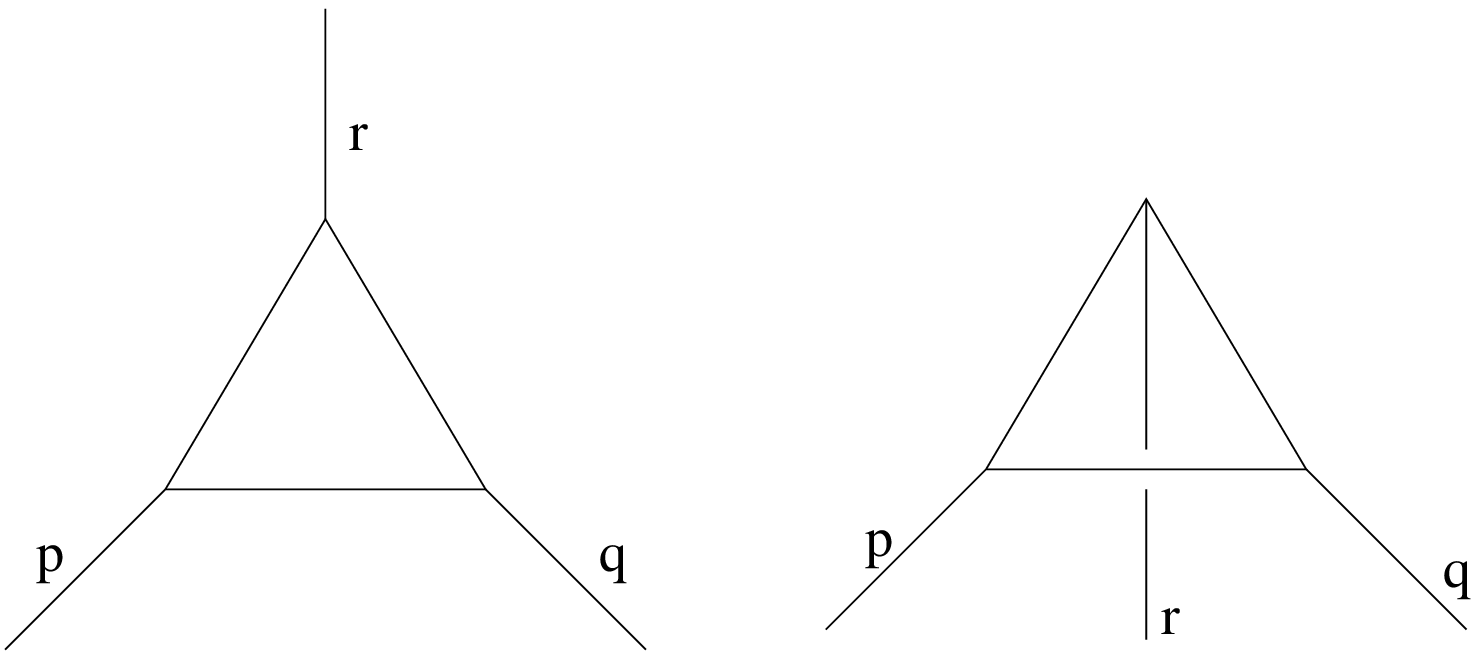}{3.0truein}
\noindent In this approximation, the nonplanar graph shown in Fig.\ 6, 
in 
terms of Schwinger parameters, gives
\eqn\vert{{1\over 2^6 \pi^3} \int {d \a d \b d \g \over (\a + \b +
\g)^3} e^{- {r \circ r +{1\over  \L^2} \over \a + \b +\g}
-m^2(\a+\b+\g) } .}
The planar graph is given by the same expression, but with $\T$ set to
zero.  The logarithmic divergence of the planar graph is effectively
cutoff in the nonplanar graph by $\L_{eff}^2(r) ={1\over r \circ r
+{1\over \L^2}}$.  Summing up the contribution of the graphs, in the
limit $\L \r \infty$, we find 
\eqn\gts{\Gamma^{(3)} = g + {g^3 \over 2^9 \pi^3} \left\{ ln \left(
{\Lambda^2  \over m^2 } \right) + ln \left({1 \over m^2 p \circ
p}\right) + \ln\left({1 \over m^2 q \circ q}\right) + ln\left({1 \over
m^2 r \circ  r}\right) \right\} + \dots}

\bigskip\bigskip\bigskip\bigskip
\cl{\it Noncommutative $\ph^4$ in four dimensions}

The graphs that contribute to the 1-loop renormalization of the
coupling constant in $\ph^4$ theory are shown schematically in Fig.\ 7.
\fig{Planar and nonplanar one loop corrections to $\Gamma^{(4)}$ in
$\phi^4$ theory.}{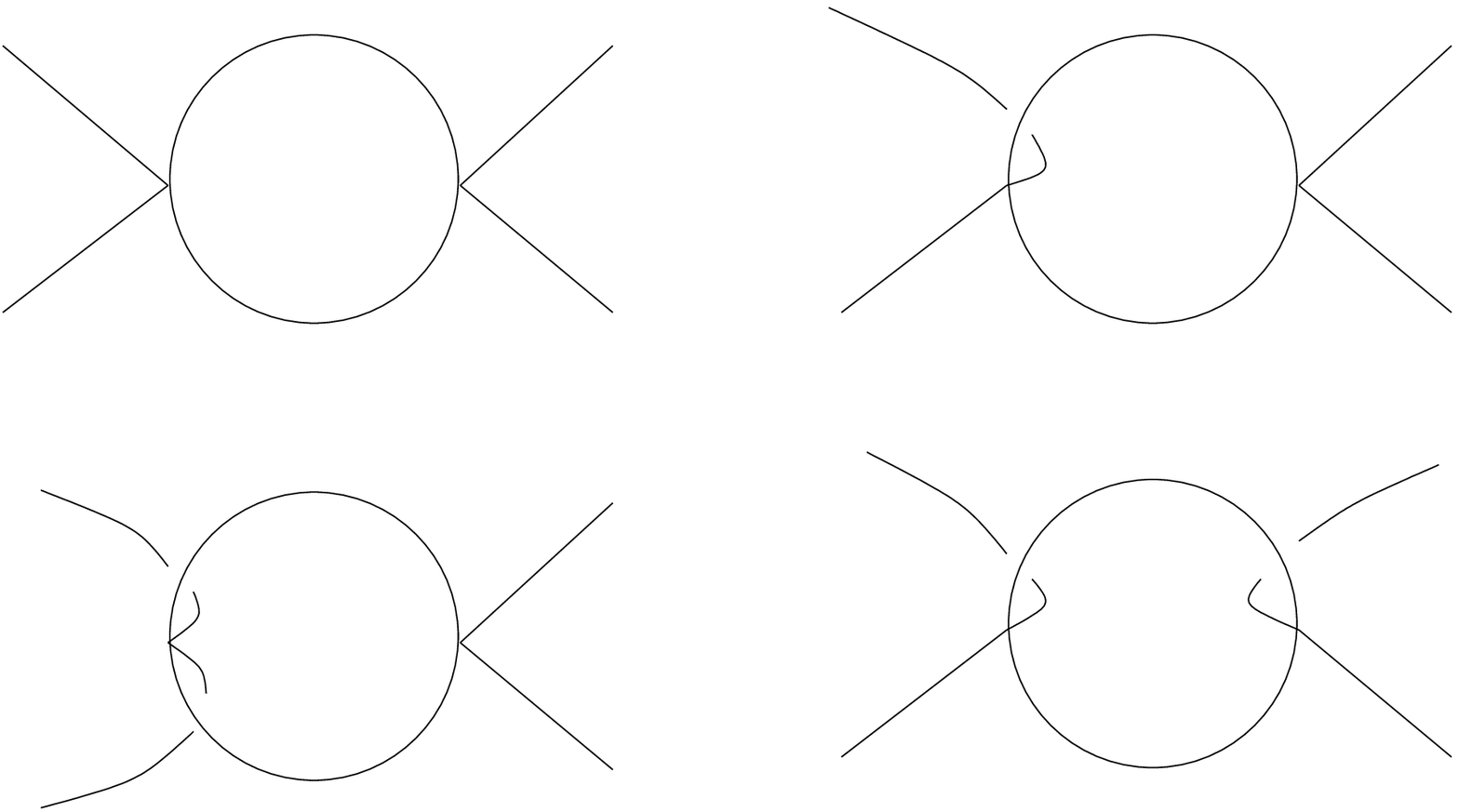}{3.0truein}
\noindent
For small external momenta we find
\eqn\etf{\eqalign{
\Gamma^{(4)} = g^2  - {g^4 \over 3 \cdot 2^5 \; \pi^2} & \left\{ 2 
ln\left({\Lambda^2 \over m^2 }\right) + ln\left({1 \over m^2 p \circ 
p}\right) + ln\left({1 \over m^2 q \circ q}\right) + ln\left({1 \over 
m^2 r \circ r}\right) \right. \cr &+ ln\left({1 \over m^2 s \circ s} 
\right) +  ln\left({1 \over m^2 (q+r)\circ(q+r)}\right) \cr &+ \left. 
ln\left({1 \over m^2 (q+s)\circ(q+s)}\right) + ln\left({1 \over m^2 
(r+s) \circ (r+s)} \right) \right\} + \dots \cr}}
For both $\phi^4$ and $\phi^3$ theories, each nonplanar graph has an 
effective cutoff $1 / P \circ P$
where $P$ is the combination of external momenta which crosses the loop.

\newsec{A Stringy Analogue}

Consider the nonplanar one loop mass correction diagram for $\ph^3$
theory, shown in Fig.\ 5. This diagram may be redrawn in double line
notation, as in Fig.\ 8. Of course, in the noncommutative field theory,
Fig.\ 8 represents a {\it particle} running in the loop and the double
line has no thickness.
\fig{The nonplanar mass correction graph for $\ph^3$ theory redrawn 
in double line notation.}{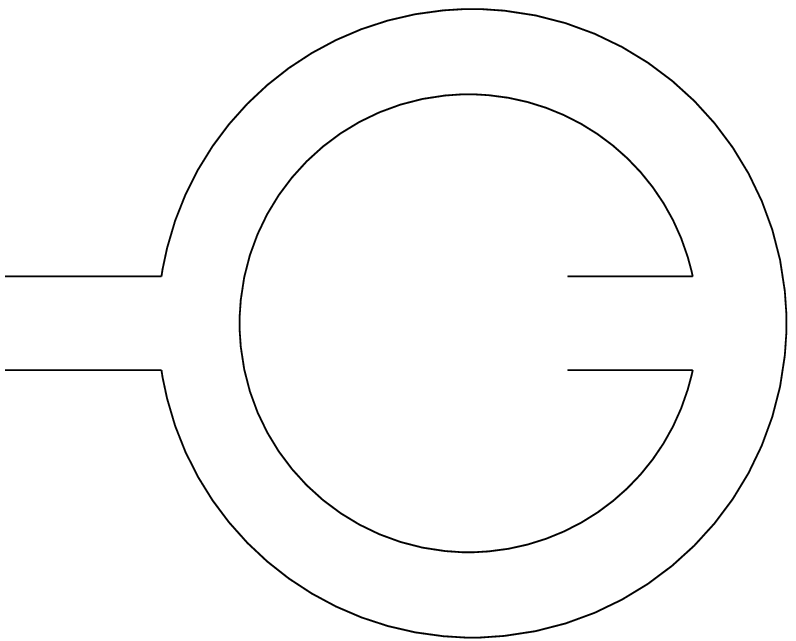}{1.7truein}

Consider, however, the 1-loop open string diagram (called the double
twist diagram) that may also be drawn as in Fig.\ 8.  In string
perturbation theory we integrate over the moduli of the diagrams. The
region of moduli space that corresponds to high energies in the open
string loop describes also the tree level exchange of a closed string
state as in Fig.\ 9.  It leads to a singularity proportional to
${1\over p_{\m} g^{\m\n}p_{\n} }$, ($g$ is the closed string metric).
UV in the open string channel corresponds to IR in the closed string
channel.

\fig{The double twist diagram in the closed string
channel.}{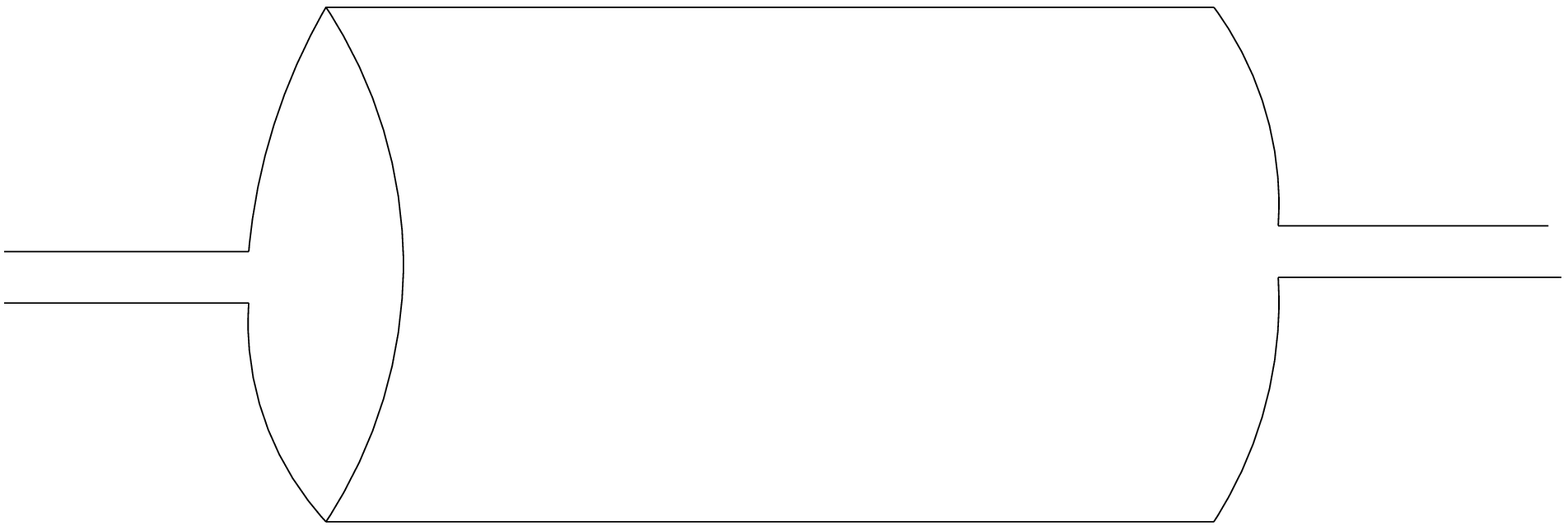}{3.0truein} 

This matches exactly with the behavior of the field theory diagram of
Fig.\ 8, if we identify the closed string metric as being proportional
to $-\T^2_{\m\n}$.  In fact this identification is very natural.  In
\rsw, it was emphasized that the analysis of D-branes in the presence
of a $B$ field involves two metrics that are related to each other.
The noncommutative gauge theory propagating on the brane ``sees'' the
open string metric.  The closed string metric governs the propagation
of closed string states in the bulk.  When the open string metric is
the identity (as in our paper), the closed string metric reduces (in
the decoupling limit) to $-\a'\T^2_{\m\n}$!

This close analogy between a one loop effect in a rather ordinary
looking (though nonlocal) field theory and an open string theory is
quite surprising, and is not completely understood.  It is an
indication of the stringy behavior of noncommutative field
theories. Other hints of the stringy nature of noncommutative theories
are the T-duality symmetry of noncommutative gauge theories on tori,
and the dominance of planar graphs in a perturbative expansion at high
momentum.  We will elaborate on these in section 7.
 
\newsec{Higher Order Diagrams}

In this section we examine the effects of noncommutativity in higher
order nonplanar diagrams. In subsection 5.1 we find examples of
nonplanar diagrams (in scalar field theories) that diverge, and
interpret these as IR divergences. In subsection 5.2 we illustrate
various other effects of noncommutativity at higher orders in a two
loop example. In subsection 5.3 we present the expression for the
Feynman integral of an arbitrary noncommutative diagram in the
Schwinger parameter representation, and comment on the structural
features of this formula. Finally in subsection 5.4, we point out the
stringy nature of `maximally noncommutative' theories obtained by
taking $\T \to \infty$.

\subsec{ Persistence of divergences in some nonplanar graphs}

\fig{Divergent higher loop graphs in $\phi^4$ theory.}
{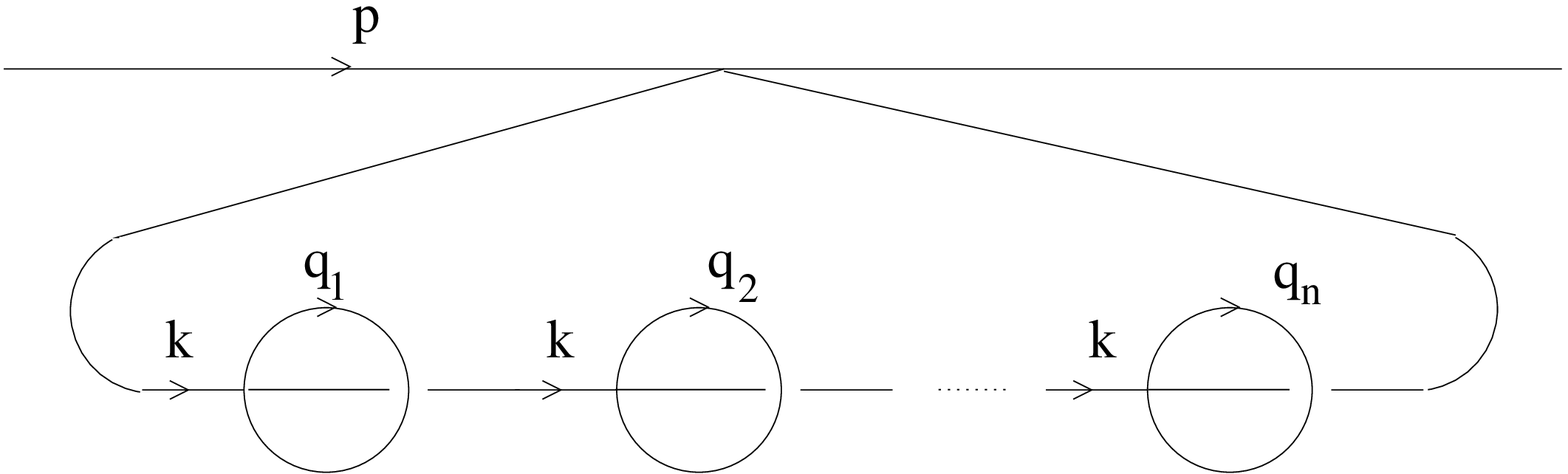}{3.0truein}

\fig{An equivalent way of representing the Feynman integral for the
graph of Fig.\ 10. The mass like vertex $v(k)$ for this graph has been
computed in section 3.}  {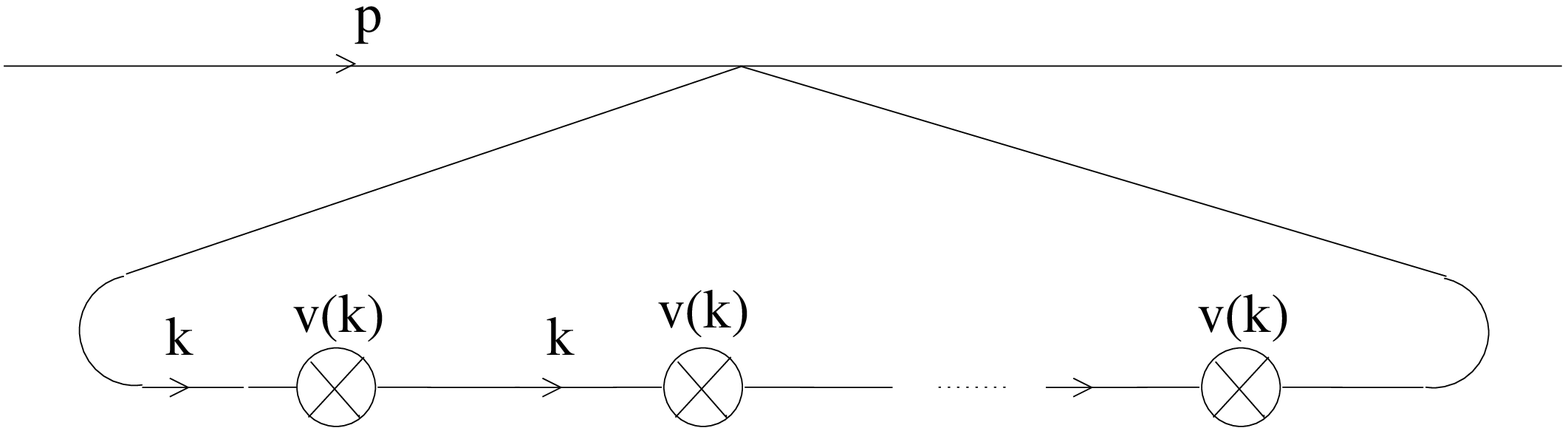}{3.0truein}

Consider the diagram $G$ of Fig.\ 10 in noncommutative $\ph^4$ theory
in four dimensions.  This graph has $n$ insertions of the nonplanar
one loop mass correction that we have computed in \plancut, and so
may equivalently be represented as in Fig.\ 11, where
\eqn\ins{v(k) = -{g^2\over 96\pi^2}{1 \over k \circ k
+{1\over \L^2}} + \ less \ singular .}
The Feynman integral for the diagram is 
\eqn\diverges{\CI(G)={1 \over (2\pi)^4} \int d^4 k {[v(k)]^n  \over
(m^2+k^2)^{n+1}}}
The contribution to $\CI(G)$ from the infrared singularities 
of $v(k)$ is proportional to 
\eqn\div{\int {d^r k_{nc}  \over (k\circ k +{1\over \L^2})^n}
\propto \cases{
\L^{2n-r}& $ 2n \geq r$ \cr 
const &$ 2n < r$ }}
($k_{nc}$ is the projection of $k$ on the noncommuting subspace in the
case where the rank $r$ of $\T$ is not maximal).  As $\L \rightarrow
\infty$, $\CI(G)$ diverges for $n \ge [ r/2]$.  This divergence occurs
at small $k_{nc}$, and high $q_i$ in the loops of Fig.\ 10.  It is a
combination of UV and IR divergence.  The presence of an
IR divergence in a massive theory is surprising, 
but is in accord with the other
IR phenomena we have seen earlier.

If we first integrate over $k$, the divergence in $\CI(G)$ appears
like a UV divergence in the integrals over $q_i$.  If, however, we
first integrate over $q_i$, the divergence appears like an IR
divergence in the integral over $k$.  We prefer the latter
interpretation, and propose to deal with the divergence in $\CI(G)$ in
a fashion similar to the standard treatment of IR divergences.

We postpone the integral over $k$ and first sum over $n$ to shift the
location of the pole in the propagator to be outside the integration
region.    More specifically,
summing the infinite series of increasingly divergent graphs of
Fig.\ 10 for all $n$ as in Fig 12, (and also including planar one loop
mass corrections) yields
\eqn\resumma{\CI= \int {d^4 k \over \G^{(2)}(k)},}
where $\G^{(2)}(k)$ is derived from \effectiveactionb\ and is given by
\eqn\rp{\Gamma^2(k) = M^2+k^2 + {g^2\over 96\pi^2 k \circ k} + \dots
\; .} 
As a result, the integral over $k$ in \resumma\ is finite (up to a
standard UV divergence).

\fig{An infinite series of divergent graphs sums up to a single
graph with the dressed one loop propagator.}
{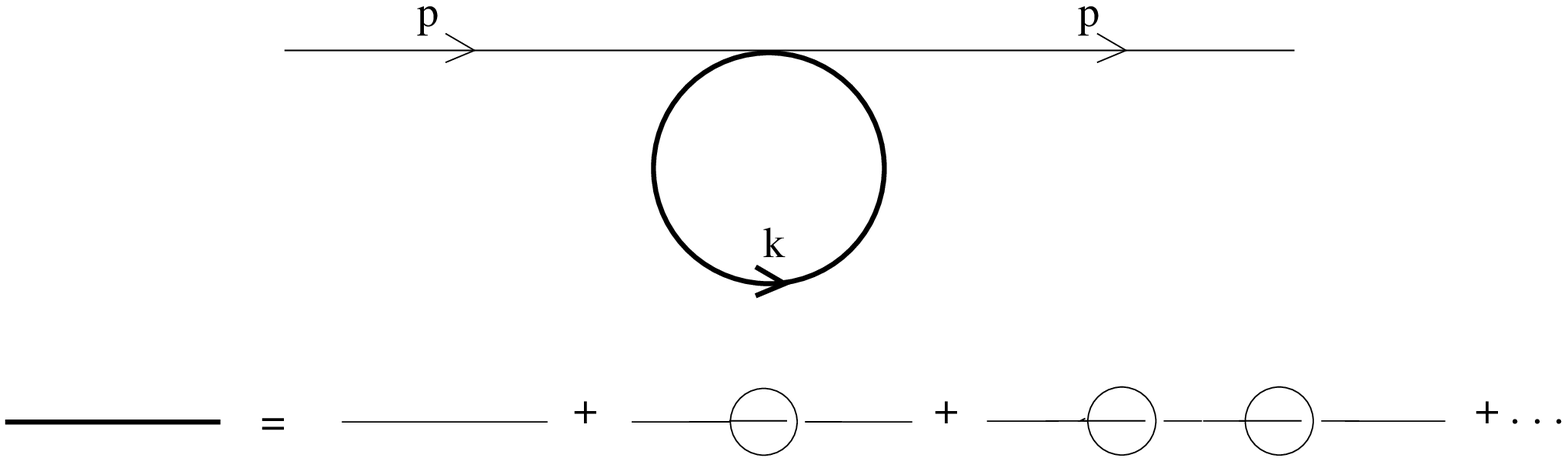}{5.0truein}

This procedure cannot be used in theories with negative $h$ because
the dangerous pole in the propagator is shifted to become tachyonic.
Then, the proper procedure is to compute the low energy effective
action of the $\chi $ field and to find its minimum.  We will not
attempt to perform such a calculation here.

We suggest that UV divergences occur only in planar graphs and all
other divergences can be treated as IR divergences, yielding finite,
physically sensible answers.  We have not proved this assertion.  It
will be very interesting to have an explicit proof of this fact.

\subsec{Other effects at two loops}

In this subsection we examine all two loop graphs that contribute
to the self mass of (for simplicity) $\ph^3$ theory in six dimensions.
This will illustrate further effect of noncommutativity present in 
higher order graphs.

Noncommutative $\ph^3$ theory in six dimensions has two classes of
mass correction graphs at two loops, corresponding to the two graphs
in the ordinary $\ph^3$ theory shown in Fig.\ 13.
\fig{The two loop contributions to the self mass of commutative
$\ph^3$ theory in six dimensions.}  {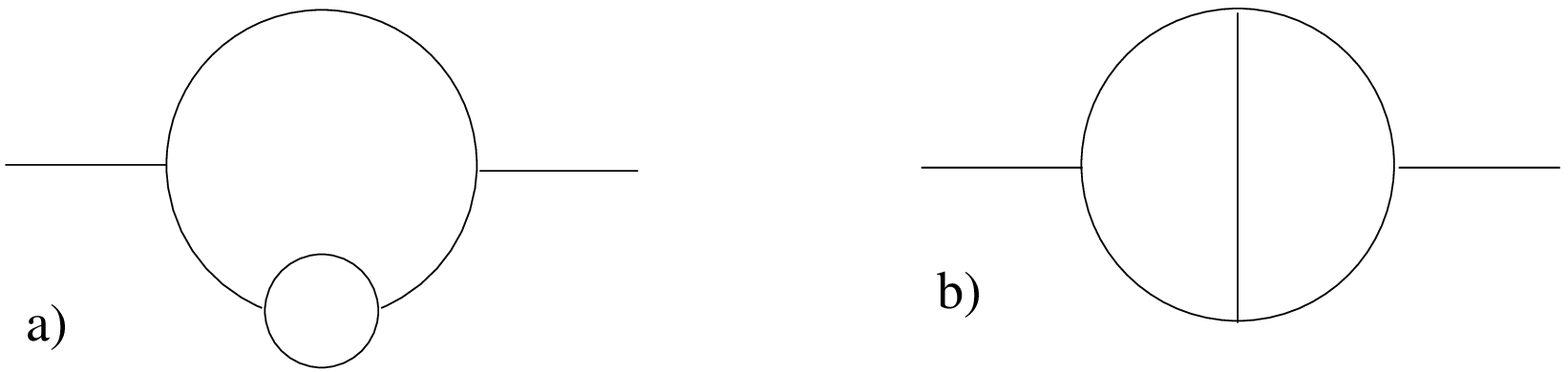}{3.0truein} 
Diagrams in the first class reduce to the graph of Fig.\ 13a at
$\T=0$.  These diagrams are members of an infinite set of graphs that
correct the propagators in the graphs of Fig.\ 5, as in the previous
subsection. The analysis of these graphs parallels that given in the
previous section for the $\ph^4$ theory.

\fig{Representatives of the three categories of graphs in
noncommutative $\ph^3$ theory that reduce to the diagram of Fig.\ 13b
at $\T=0$.} {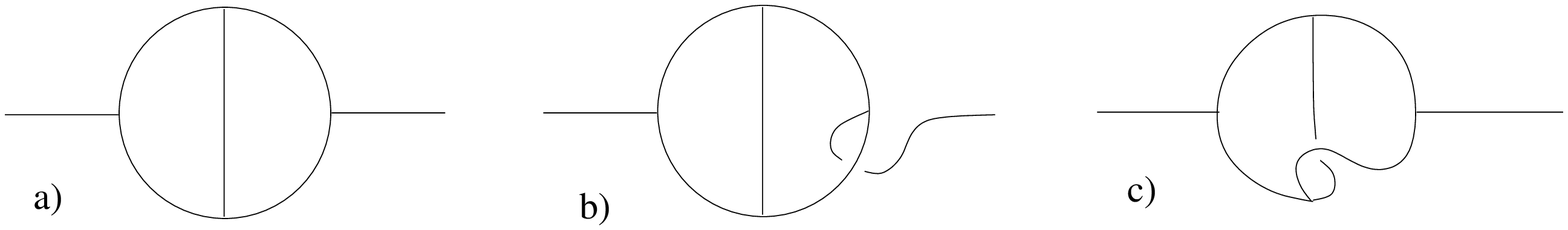}{5.0truein}

The diagrams that reduce to the graph of Fig.\ 13b at $\T=0$ fall into
three categories, representatives of which are shown in Fig.\ 14.
Fig.\ 14a shows the single planar graph. The corresponding Feynman
integral does not depend on $\T$, and evaluates to ${1 \over 8}$ times
the result for the graph of Fig.\ 13b in the commutative theory. As in
the commutative theory, this graph contains overlapping
divergences. These are dealt with in the usual way, by combining it with
the counterterm graphs of Fig.\ 15a and Fig.\ 15b, whose contributions
are also suppressed by a factor ${1 \over 8}$ relative to the
commutative theory.  Thus, as in the commutative theory, the graphs of
Figs.\ 14a, 15a and 15b combine to give a result whose divergence may
be cancelled by counterterms.
\fig{One loop mass correction graphs with one loop vertex correction 
counterterms.} {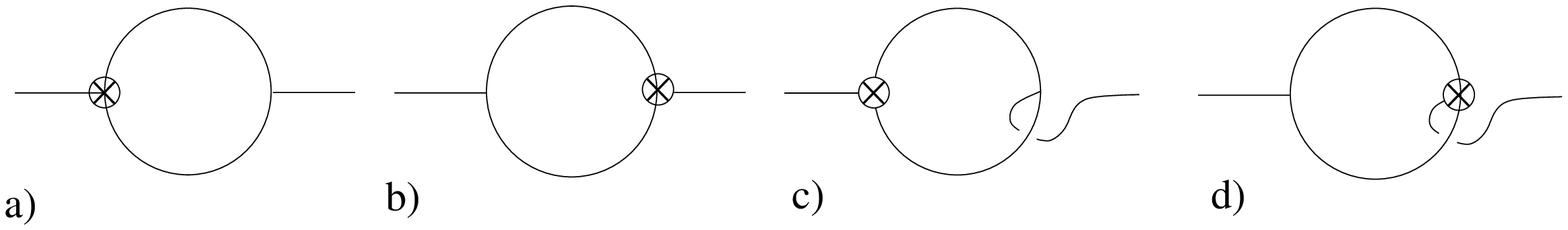}{5.0truein}

Fig.\ 14b represents a set of three nonplanar graphs, each of which is
nonplanar only because an external line overlaps with an internal
line.  Two of these graphs contain divergent planar vertex correction
subgraphs and should respectively be combined with the one loop graphs
containing counterterms shown in Figs.\ 15c and 15d.  Once their
subdiagrams have been renormalized, these diagrams, as well as the
third diagram in this set, are finite, effectively cut off by
$\L_{eff}={1 \over \sqrt{p \circ p}}$.  At large $\L_{eff}$, these
graphs are all proportional to $ g^4\L_{eff}^2 \ln(\L_{eff})$,
providing a logarithmic correction to the leading singularity in
\efff.  The proportionality of these diagrams to $\L_{eff}^2
\ln(\L_{eff})$ (rather than $\L_{eff}^2$) is a consequence of the fact
that fewer counterterms are added in the noncommutative theory than
in the commutative theory.

The four remaining two loop mass correction graphs, which include the
diagram of Fig.\ 14c, are nonplanar due to the crossing of internal
lines. As we will show in the next subsection, these graphs are
finite, effectively cutoff by $1/\sqrt{\theta}$ where $\theta$ is the
largest eigenvalue of $\Theta^{\mu \nu}$. In contrast to the nonplanar
one loop graphs and the graphs considered in the previous paragraph,
these diagrams are nonsingular at low momenta, since the effective
cutoff is independent of external momenta. This is a consequence of
the fact that the noncommutative phase factor responsible for removing
the divergence depends only on internal momenta, which are integrated
over.

\subsec{General Higher order graphs}

In the appendix, we express the Feynman integral for an arbitrary graph
($G$) in terms of an integral over Schwinger parameters. Up to overall 
constants, the result
(part of which has been independently reported in \riouri) is 
\eqn\resu{\delta(\sum p_I) V(p^I) \int d \alpha_i e^{-\alpha_i m_i^2} {1 
\over
(\prod_{\theta_a} P_G(\alpha, \theta_a))^{1 \over 2}} e^{-p^I_\mu
\left(  {S_{IJ}(\alpha,\Theta^2) \over P_G(\alpha,\Theta^2)} 
\right)^{\mu \nu}p^J_\nu  - i p^I_\mu  \left( {\Theta
A_{IJ}(\alpha,\Theta^2) \over P_G(\alpha,\Theta^2)} \right)^{\mu
\nu}p^J_\nu}.}
Here $P$, $S$, and $A$ are homogeneous polynomials in $\{\alpha_i,
\Theta\}$, described in terms of graph theoretic properties of $G$ in
the Appendix, the phase $V(p^I)$ is as in \phasefa, and $\theta_a$ are
the eigenvalues of $\Theta$.

Four terms in this expression involve the noncommutativity parameter
$\Theta$:
\item{1)} The first factor $V(p^I)$ is the overall phase, present for
any graph.  It depends on the cyclic order of the external momenta
$p$.  This is the sole effect of non-commutativity for planar graphs.
\item{2)} The denominator $P_G$ is a homogeneous polynomial in
$\{\alpha_i, \Theta \}$ (independent of external momenta) of degree
$L$, where $L$ is the number of loops in $G$. Divergences of a graph
are associated with regions of parameter space where $P_G \rightarrow
0$.  Crossing internal lines of $G$ give rise to $\T$ dependent terms
in $P$, and these reduce the rate at which $P$ becomes small when the
$\alpha$'s are scaled to zero. More precisely, the degree of $P_G$ as
a function of $\T$ is $2h$, where $h$ is the genus of $G$ (with
external lines removed), so a graph with superficial degree of
divergence $\omega$ at $\Theta=0$ will essentially have a superficial
degree of divergence $\omega - 2hr$ for $\Theta$ of rank $r$. Thus,
crossing internal lines regulate the divergences with an effective
cutoff of order ${1\over \sqrt{\T}}$ (assume for simplicity that $\T$
has maximal rank and all its entries are of the same order of
magnitude), as for the graph of Fig.\ 14c.  In particular, this
effective UV cutoff depends only on $\T$ and not on the momenta.
\item{3)} The exponential factor $exp(- p_\mu^I(S_{IJ}/ P_G)^{\mu \nu}
p_\nu^J )$ acts to reduce divergences when an external momentum $p$
crosses an internal line.  If the momenta associated with some
subgraph $G_q$ of $G$ crossed by $p$ are scaled by $\rho$, then the
exponential scales as $exp(-p \circ p / \rho)$. Thus the exponential
cuts off possible divergences associated with $G_q$, with an
effective cutoff ${1 \over \sqrt{ p \circ p}}$, as we have seen for
the one loop graphs of section 3.  As there, this cutoff depends both
on $\T$ and on the momenta and becomes large as $p \circ p \to 0$, 
resulting in the singular IR behavior we have seen previously.
\item{4)} The final exponential factor in \resu\ is a phase in the
integrand which depends on the Schwinger parameters.  This modifies
the behavior of finite graphs, but does not seem to affect the
convergence of the graph.

\noindent
Using these expressions, it is possible to demonstrate the convergence
(at $\T \neq 0$) of the Feynman integral associated with $G$, if $G$
has no divergent planar subgraphs and $\o(G_i) \leq 0$, for all
subgraphs $G_i$ of $G$, where $\o$ denotes the degree of divergence of
a graph.

As we have remarked above, the issue of renormalizability of scalar
noncommutative field theories is quite subtle.  Some nonplanar graphs
are divergent, but we have suggested that the divergence should be
viewed as an IR divergence and should be handled appropriately.
Therefore, we think it will be quite interesting to prove the
renormalizability of these theories to all orders.

\subsec{Maximal noncommutativity}

It is interesting to consider the limit in which we take the
noncommutativity parameter $\T$ to infinity.  In this limit the theory
is maximally noncommutative.  Equivalently, we can hold $\T$ fixed and
scale all the momenta (and masses) to infinity.  If the theory had
been an ordinary field theory, and it had not had any mass parameters,
this would have been the ``short distance limit.''

{}From \resu\ it may be seen that in this limit the amplitudes (at 
generic values of external momenta) are dominated by the planar graphs.  
For nonplanar graphs in which internal lines cross, the polynomial $P_G$
blows up as $\T \to \infty$. For nonplanar graphs in which external
lines cross internal lines, the factor $S_{IJ}/P_G$ behaves like a
positive power of $\Theta$, so the exponential tends to zero as
$\Theta \to \infty$ (for generical external momenta). Only planar
graphs, which have no $\Theta$ dependence apart from the overall
phase, survive the $\Theta \to \infty$ limit.

This observation is extremely interesting.  The maximally
noncommutative theory, which is obtained in the limit $\T\to \infty$
is given by the planar diagrams only.  If $\phi$ is an $N\times N$
matrix the theory in this limit is independent of $N$.  In particular,
it is the same as the large $N$ theory.  We therefore conclude that in
the high momentum or maximal noncommutativity limit, the diagrams
become planar and the theory appears to be stringy.

In Section 4 we pointed out the close similarity between the singular
IR effects in the nonplanar one loop diagram here and in open string
theory.  The T-duality of noncommutative Yang-Mills theories on a
torus is yet another indication of the stringy nature of
noncommutative field theories.  Finally, in the large $\T$ limit the
theory becomes a string theory.  We find it intriguing that these
field theories exhibit so many stringy phenomena.

\newsec{Origin of UV/IR Mixing}

In this section we will give some intuitive explanations for why the
noncommutativity of spacetime leads to the surprising mixture of UV
and IR. Roughly, very small pulses instantaneously
spread out very far upon
interacting. In this manner very high energy processes have important
long distance consequences.

\subsec{Nonlocality of the star product}

We may rewrite the star product \starp\ of two functions in position
space as \rfilk\
\eqn\convolution{\big(\ph_1 \star \ph_2 \big) (z) = \int d^dz_1 d^dz_2
\ph_1(z_1) \ph_2(z_2) K(z_1, z_2, z) }
with the kernel
\eqn\ker{K(z_1, z_2, z) = {1\over \det(\Theta)}
e^{2i(z-z_1)^{\m}\Theta^{-1}_{\m\n}(z-z_2)^{\n}}.}
Since $|K(z_1,z_2,z)|$ is a constant independent of $z_1, z_2, z$, the
star product appears to be infinitely nonlocal.  However, 
the oscillations in the phase of $K$ suppress parts of the integration
region.  We now make this statement more precise. 

Consider for simplicity the case $d=2$ with coordinates $[x,y] = i\t$. 
Let $\phi_1$ be a function which has widths of order $\D_{x1}$ and
$\D_{y1}$ in the $x$ and $y$ directions, and assume that it varies
slowly over its domain of support.  Then, the integral over $x_1$ in
\convolution\ is proportional to
\eqn\loc{\int d x_1  \phi_1(x_1, y_1) e^{i x_1 (y_2 - y) / \t}.}
It is suppressed by phase oscillations if
\eqn\sca{\Delta_{x1} |y_2 - y| / \t \gg 1.}
We conclude that $(\phi_1 \star \phi_2)(x,y)$ samples $\phi_2$ in a
width $\d y_2 \approx  {\t \over \D_{1x}}$ about $y$.

In a similar way we arrive at the estimates
\eqn\nonlocal{\d y_2 \D_{1x} \approx \t, \; \; \;
\d y_1 \D_{2x} \approx \t, \; \; \;
\d x_2 \D_{1y} \approx \t, \; \; \;
\d x_1 \D_{2y} \approx \t, }
where $\D_{ai}$ is the width of $\ph_a$ in the $i$ direction and
$\d_{ai}$ is an estimate of the nonlocal width of the $a^{th}$
function sampled by the star product in the $i^{th}$ direction. This
is illustrated for Gaussian wave packets in Fig.\ 16.
\fig{Star product of Gaussian wave packets $\phi_1$ and $\phi_2$.
If $\ph_1$ and $\ph_2$ are nonzero only in the shaded regions of the
diagram, $\left( \ph_1 \star \ph_2 \right)$ is nonzero in the
intersection of the dotted regions of the diagram. The dotted regions
are constructed as described in the text.}  {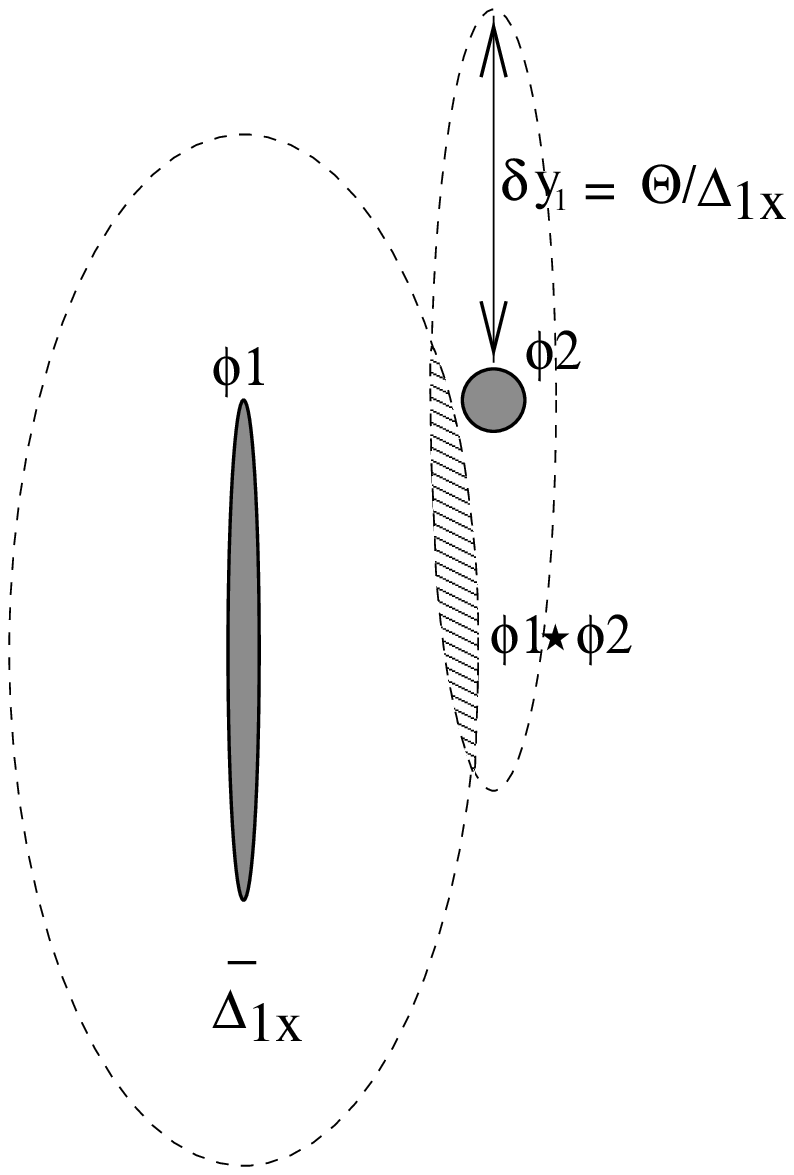}{2.0truein}

For the important special case $\ph_1=\ph_2=\ph$, where $\ph$ is
smooth and of width $\D_{x}, \D_{y}$, the widths $\delta_x$ and 
$\delta_y$ of $\ph \star \ph$ are given by
\eqn\mpm{\d_x \approx max( \D_x ,{\t \over \D_y}), \  \ \d_y \approx
\max(\D_y,{\t \over \D_y}).}
Thus if $\ph$ is nonzero over a very small region of size $\D \ll 
\sqrt{\t}$, $\ph \star \ph$ is nonzero over a much larger region
of size ${\t  \over \D}$.  This nonlocality has important consequences
for the dynamics. 

\subsec{Consequences for Dynamics}

Consider, for example, noncommutative $\ph^3$ theory \phicuac.
Classically, $\ph(x)$ obeys the equation
\eqn\pert{\left( \p^2 -m^2 \right) \ph(x) 
={g\over 2} \left(\ph \star \ph\right)(x).}
Given a solution $\ph_0(x)$ to the free equation of motion, the
corresponding solution to \pert\ is given perturbatively by
\eqn\pho{\ph(x)=\ph_0(x)-{g\over 2} \int d^dy G(x-y) \left(
\ph_0 \star \ph_0 \right)(y)+... }
where $G(x)$ is the appropriate Green's function.  At first order the
effective source is ${-g \over 2}\left( \ph \star \ph \right)(x)$.  

As demonstrated in the previous subsection, if $\ph_0(x)$ is nonzero
over a region of size $\D$, $\left( \ph_0 \star \ph_0 \right) (x)$ is
nonzero over a region of size $\d= \max(\D , {\t \over \D})$ ($\t$ is
a typical entry of the matrix $\T$).  $\d$ is very big when $\D$ is
very small.  Thus, classically, pulses of size $\D \ll \sqrt{\t}$
spread to size $\d={\t \over \D}\gg \sqrt{\t} $ upon interacting. The
extent of the spread is independent of the mass of the particle.

In the quantum theory, even very low energy processes receive
contributions from high energy virtual particles.  In a nonplanar
graph, a virtual particle of energy $\o \gg {1 \over \sqrt{\t}}$ (size 
${1\over
\o}\ll \sqrt{\t}$) will, upon interacting, spread to size $\o \t$,
producing important effects at energy ${1\over \t \o}$.  Therefore,
imposing a UV cutoff $\L$ on $\o$ effectively imposes an IR cutoff
$\L_{IR}={ 1 \over \t \L}$ on IR singularities produced by the
nonplanar graph. Notice that the IR effects produced by high energies
in loops are independent of the mass of the particle.  This gives us a
qualitative explanation for the effects observed in section 3.

\newsec{Noncommutative Yang-Mills}

It is difficult to repeat the analysis of the previous sections for 
gauge theories for several reasons. 
\item{a)} Noncommutative gauge theories have no local gauge invariant
operators. The most easily constructed gauge invariant operators, (for
example $\int d^dx Tr \hat F^2)$ involve an integration over all
space.  Approximately local gauge invariant operators may be
constructed as follows.  Using the change of variables from the
noncommutative gauge fields $\hat A$ to the commutative gauge fields
$A$ \rsw\ we can construct local gauge invariant operators in terms of
$A$, such as $\Tr F^2$, and express them in terms of the
noncommutative variables $\hat A$.  These are manifestly gauge
invariant, however they are not local.  They are given by a power
series in $\T$ and are therefore ``smeared'' over a region of size of
order the only scale in the problem, i.e.\ $\sqrt \T$ (again, we
assume for simplicity that $\T$ has maximal rank and all its entries
are of the same order of magnitude)\foot{These gauge invariant
operators have been suggested by several people among them D. Gross,
S. Kachru and E. Silverstein.}.  Although such gauge invariant
observables exist, it is not clear how to calculate with them and what
their properties are.  Therefore, it is not clear to what extent the
IR effects we encountered above persist in these theories.
\item{b)}  Noncommutative Yang-Mills theory has massless fields.  Thus
it is difficult to disentangle the usual IR effects from possible IR
singularities induced by the noncommutativity.
\item{c)} At low energies, noncommutative gauge theories are strongly
coupled in terms of the elementary quanta $\hat{A}$, and it is not
clear what the light degrees of freedom are.  Therefore, a low energy
effective theory like the one we analyzed in the scalar field theory
is not easy to construct.
\item{d)} Some of the effects observed in section 3 (for example, 
the new pole) resulted from the existence of quadratic divergences. 
Such divergences are absent in Yang-Mills theory which has at worst
logarithmic divergences.  The logarithmic divergences do lead to IR
singularities but they are not as pronounced as the poles we observed
and discussed above.

In subsection 7.1 below we present the quadratic 
one loop 1PI effective action for
$U(N)$ noncommutative Yang-Mills.  In subsection 7.2 we determine the
energy dependence of the dilaton in the string theory dual to
noncommutative Yang-Mills, and compare this with the prediction from
the AdS/CFT correspondence.

\subsec{One Loop in Noncommutative Yang-Mills}

The Lagrangian for noncommutative $U(N)$ Yang-Mills theory is given by
\eqn\acncym{\CL={1\over 4g^2} \int \Tr( \hat F_{\m\n}\hat  F_{\m\n}),}
where $\hat A=\hat A^a t^a$ ($\Tr(t^a t^b)=\d^{ab}$) and
\eqn\fs{\hat F_{\m\n}= \p_{\m}\hat A_{\n}-\p_{\n}\hat
A_{\m}-i\left(\hat A_{\m}\star 
\hat A_{\n} -\hat A_{\n} \star \hat A_{\m}\right)}
is covariant under noncommutative gauge transformations
\eqn\gt{\d \hat A_{\m}=\p_{\m}\ep-i\left(\hat A_{\m}\star \ep -\ep
\star \hat A_{\m} \right) = \hat D_{\m} \ep.}
Let
\eqn\linf{ (\p \hat{A})_{\m\n}=\p_{\m}\hat A_{\n}-\p_{\n}\hat
A_{\m}.}
The divergent part of the quadratic 
one loop 1PI effective action of the
ordinary $U(N)$ in the background field gauge is given by
\eqn\pincym{{\G^1_2(\T=0)={1\over 4} \int\left( ({1\over g_0^2}
+{\b(g_0^2) \over 2g_0^4} \ln {\Lambda_0^2 \over {k^2}} )
\Tr(\p A)^2 -{\b(g_0^2) \ln {\Lambda_0^2 \over {k^2}}
\over 2g_0^4 N}(Tr \p A)^2 \right)}, }
where
\eqn\betasun{\b(g^2)={-11 g^4N^2\over 12 \pi^2}}
is the one loop beta function for $SU(N)$ Yang-Mills, $\Lambda_0$ is
an ultraviolet cutoff, and $g_0$ is the bare coupling at scale
$\Lambda_0$.

\fig{Nonplanar one loop $U(N)$ diagrams contribute only to the
$U(1)$ part of the theory}{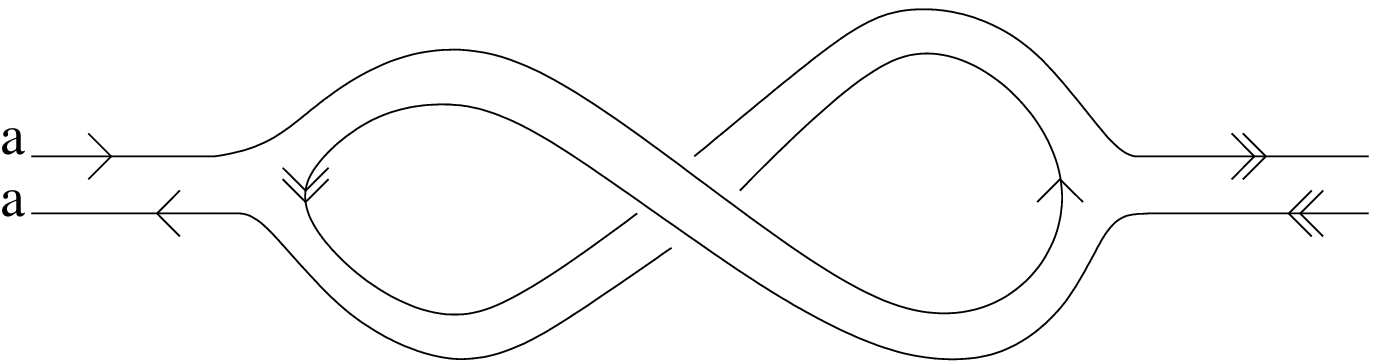}{2.5truein} 

The only nonplanar contribution to $\G^1$ is the term ${1\over 4}
\int {\b(g^2) \ln {\Lambda_0^2 \over k^2}\over 2g_0^4 N}(Tr \p A)^2$,
which receives contributions from diagrams like those of Fig.\ 17.  It
contributes only to the $U(1)$ part of the action.  In the
noncommutative theory, the divergence in this graph will be cut off by
$1/\sqrt{p \circ p}$. (Since star products and matrix products always
appear together, graphs which are non-planar in the matrix sense will
also be nonplanar in the sense we have considered in the rest of this
paper.) The divergent part of the quadratic 
noncommutative one loop effective
action is therefore given by
\eqn\ncymmopi{{\G^1_2={1\over 4} \int\left( ({1\over g_0^2}
+{\b(g_0^2) \over 2g_0^4} \ln {\Lambda_0^2 \over {k^2 }} )
\Tr( \p \hat A)^2 -{\b(g_0^2) \ln {1 \over (k\circ k){k^2}}
\over 2g_0^4 N}(\Tr \p \hat A)^2 \right)}}
After renormalization, 
\eqn\opinnn{\G^1_2={1\over 4} \int\left( ({1\over g^2(\L)}
-{11N\over 24 \pi^2} \ln {\Lambda^2 \over {k^2}} )
\Tr( \p \hat A )^2 +{11 \over 24 \pi^2}\ln {1\over k^2 (k \circ k)}
(\Tr \p \hat A )^2 \right)}
where $\L$ is an arbitrary finite energy scale, and $g^2(\Lambda)$ 
is the solution to the equation
\eqn\gevol{ {dg^2 \over d \ln \Lambda}=\b(g^2) }
with initial condition $g(\L_0)=g_0$.

We make two comments:
\item{a)} The one loop beta function for noncommutative $U(N)$ Yang
Mills is given for all $N$, including $N=1$, by \betasun.  Note, in
particular, that the $N=1$ theory is also asymptotically free.  The
computations in \refs{\rjabbari-\rruiz} confirm \betasun\ for $N=1$.
We stress that the beta function for the noncommutative theory can be
found without doing any new calculation by simply using the known
value of the beta function of the ordinary commutative $SU(N)$ theory.
\item{b)} The $U(1)$ part of the quadratic effective action has a
logarithmic IR singular piece.  This is similar to the the
wavefunction `renormalization' in \efff. Note that, as for the scalar
theory, this one loop logarithm is significantly corrected by higher
order effects at energies for which it is important.

\subsec{Energy Dependence of the Dilaton}

As we have argued in section 5.4, noncommutative theories are
dominated by planar graphs,
 and therefore seem stringy as $\Theta \to
\infty$. Specifically, by the formulae of section 5.3, a 
noncommutative Feynman diagram of genus $h$ (ignoring external lines) is 
suppressed
 relative to a planar
graph by the factor of
\eqn\dilat{{1 \over N^{2h}}\prod_{i=1}^{{r\over 2}}{1\over 
(E^2\t_i)^{2h}},}
where $r$ is the rank of $\Theta$, whose eigenvalues are $\{ \t_1,
-\t_1..., -\t_r \}$. The contribution of genus $h$ graphs to the
$U(N)$ noncommutative gauge theory is expected to reflect the
contribution of genus $h$ worldsheets of the `noncommutative 
QCD string', whose dilaton $\ph$ must therefore scale with energy 
according to 
\eqn\dilato{e^{-\ph(E)} \propto {1 \over N}\prod_{i=1}^{{r\over 
2}}{1\over E^2\t_i}.}

\nref\rmr{J. Maldacena and J. Russo, ``Large N Limit of Non-Commutative
Gauge Theories,'' hep-th/9908134.}%
\nref\rih{ A. Hashimoto and N. Itzhaki, ``Non-Commutative Yang-Mills
and the AdS/CFT Correspondence,''
hep-th/9907166.}%
\nref\rssk{S. Das, S. Kalyana Rama, S. Trivedi, 
``Supergravity with Self-dual B fields and 
Instantons in Noncommutative Gauge Theory,''hep-th/9911137.}%

This prediction may be tested\foot{See also \rishibashi.}
 for the $\CN=4$ noncommutative $U(N)$
theory. The string theory dual to $\CN=4$ noncommutative $U(N)$ is IIB
theory on the spacetime background described in \refs{\rmr - \rssk}.
Fields in this background depend on the radial coordinate $u$ ($u=0$
at the horizon, $u=\infty$ at the boundary), which is identified with
the energy scale in the dual field theory, via the UV-IR
correspondence.  The dilaton dependence in these solutions is indeed
\eqn\dilat{e^{-\ph(u)} \propto {1 \over N} \prod_{i=1}^{{r\over
2}}{1\over u^2\t_i}} 
at large $u$.

\bigskip
\centerline{\bf Acknowledgements}

We would like to thank S. Chakravarty for collaboration at the initial
stages of this project, several useful discussions and helpful
comments.  We thank O. Aharony, T. Banks, M. Douglas, R. Gopakumar,
D. Kutasov, H. Liu, J. Maldacena, M. Rangamani, E. Rabinovici,
A. Schwimmer, E. Silverstein, W. Taylor, E. Witten and
A.B. Zamolodchikov for useful discussions.  S.M. would like to thank
the Rutgers string group for hospitality and several stimulating
discussions.  The work of S.M. was supported in part by DOE grant
\#DE-FG02-91ER40671, and N.S. by grant
\#DE-FG02-90ER40542.

\appendix{A}{Feynman integral for a general graph in a noncommutative
theory}

Here we present an expression for the Feynman integral corresponding
to an arbitrary graph in a noncommutative theory, written as an
integral over Schwinger parameters corresponding to each internal
line. This expression reveals several generic effects of the
non-commutativity, which are discussed in section 5.3. The derivation,
which is somewhat technical, will not be presented here.

We begin with a scalar field theory with no derivative couplings and 
consider a graph $G$ with $E$ edges, $V$ vertices, $L$ loops, and 
external momenta $p_I$ (the generalization to arbitrary theories is 
given in the next subsection). We introduce Schwinger parameters 
$\alpha_i$ for each internal line of the graph, after which the 
integrals over momenta may be evaluated exactly to give
\eqn\fd{c(g_i) \delta(\sum p^I) V(p^I) {1 \over 2^{dL} \pi^{dL/2}} \int 
d\alpha_i
e^{-\alpha_i m_i^2} \CI(\alpha, \Theta, p^I).}
Here, $c(g_i)$ is an overall constant given by the product of the 
symmetry factor for the graph with all vertex factors. The factor 
$V(p_I)$, defined in \phasefa, is
the overall phase, present for all graphs, which depends on the cyclic
ordering of $p$'s.  The integrand $\CI$ is
\eqn\integrand{\CI(\alpha, \Theta, p^I) =   {1 \over
(\prod_{\theta_a} P_G(\alpha, \theta_a))^{1 \over 2}} e^{-p^I_\mu
\left(  {S_{IJ}(\alpha,\Theta^2) \over P_G(\alpha,\Theta^2)} 
\right)^{\mu \nu}p^J_\nu  - i p^I_\mu\left( {\Theta
A_{IJ}(\alpha,\Theta^2) \over P_G(\alpha,\Theta^2)} \right)^{\mu
\nu} p^J_\nu }.}
In the rest of this subsection we explain the various objects in this
expression.  In the denominator, the product runs over the
eigenvalues $\theta_a$ of $\Theta$.  The function $P_G(\alpha,
\theta)$ is a polynomial defined by
\eqn\denom{P_G=P(G) = \sum_{H \in R(G)} \theta^{E-V+1} \prod_{i \in
H^c} {\alpha_i \over \theta}.}
Here $H^c$ is the set of internal lines in the complement of $H$ and
$R(G)$ is the set of subgraphs
$H$ of $G$ containing only internal lines such that $H$ contains a
maximal tree (a tree touching all vertices) of $G$ and $H$ has maximal
genus (a single index line when drawn in double line
notation)\foot{The condition of maximal genus may also be stated as
$\det(C_{ij}(H/T))=1$, where $C_{ij}$ is the
intersection matrix for the single vertex graph $H/T$ obtained by
starting with $H$, removing external lines, and contracting all lines
of $T$ to a single point.}.  We note that $P$ is an even polynomial in
$\Theta$, of degree $2h$, where $h$ is the genus of the graph $G$ 
(ignoring external lines).

$S$ and $A$ in \integrand\ are polynomial defined in terms of $P$ as
\eqn\defS{\eqalign{
&S_{pq}(\alpha, \theta) = {1 \over 2}( P^*(G_{p=r}) +
P^*(G_{q=r}) -  P^*(G_{p=q}))\cr
&A_{pq}(\alpha, \theta) = {1 \over \theta} [P^*(G_{p,q=r})P(G) -
P^*(G_{p=r}) P^*(G_{q=r}) \cr 
&\qquad \qquad\qquad + {1 \over 4} (P^*(G_{p=r}) + P^*(G_{q=r}) -
P^*(G_{p=q}))^2]^\half ,}} 
where $r$ is a particular external momentum (all choices give
equivalent results), $G_{p=q}$ is a graph obtained by connecting
external lines $p$ and $q$ to form a new internal line and $G_{p,q=r}$
is a graph obtained by connecting external momenta $p$ and $q$ to the
vertex connected to $r$ (without creating any new intersections). Here
$P^*$ means that $P$ is to be evaluated with Schwinger parameters for
any added lines set to zero.

\subsec{Generalization to arbitrary theories}

The result of the previous subsection may be easily generalized to
arbitrary theories which may include particles of non-zero spin or
derivative couplings.  In terms of the momentum space Feynman
integrands, these theories will have additional factors of internal
and external momenta in the numerator coming from propagators or
vertex factors. The resulting expressions for $\CI$ (the Schwinger
parameter integrand) may be related to the expressions obtained above
for non-derivative scalar theories by the following trick.

Consider an arbitrary graph $G$ in a general theory, and let $\{ l_j
\} $ be the set of internal momenta which appear (possibly more than
once) in the numerator of the Feynman integral. We define a new graph
$G^+$ as follows.  For each $l$, let $a$ be the vertex which $l$
leaves. Now include two new external momenta, a momentum $r^{l}$ which
leaves vertex $a$ immediately counterclockwise from $l$ and a momentum
$-r^l$ which leaves $a$ immediately clockwise from $l$. Since these
momenta sum to zero, they do not affect overall momentum conservation,
and in fact their only effect is to multiply the integrand with an
extra phase
\eqn\np{e^{i  r^l \times l}.}
We may now replace all internal momenta $l_\mu$ appearing in the
numerator of the Feynman integral with derivatives
$-i(\Theta^{-1})_{\mu \nu} \partial_{r^l_{\nu}}$ acting outside the
integral. The resulting integral over internal momenta is exactly the
expression $\CI(G^+)$ for the graph $G^+$ in a non-derivative scalar
theory. Thus, if the function of momenta in the numerator of the
Feynman integral for $G$ is some polynomial $\pi(l_\mu, p_\mu)$, we
have
\eqn\deriv{\CI_G(\alpha, \Theta) = \pi(-i(\Theta_0^{-1})_{\mu \nu}
\partial_{r^l_{\nu}},p_\mu) \; \CI_{G^+}(\alpha, \Theta_0) |_{r^l_\mu =
0, \Theta_0 = \Theta} .}
Here, $\CI_{G^+}$ may be explicitly evaluated using the rules of the
previous subsection for non-derivative scalar theories. In order that
this expression be well defined for $\Theta$ of any rank, we must
evaluate $\CI_{G^+}$ using an invertible $\Theta_0$ and then set
$\Theta_0 = \Theta$ at the end. Interestingly, this procedure may be
used to give Feynman integrals for a general theory in terms of
Feynman integrals for a nonderivative scalar theory even in the
commutative case of $\Theta = 0$.

\listrefs
\end